\documentclass[aps,prd,preprint,showkeys,nofootinbib,superscriptaddress]{revtex4-2}

\usepackage{tikz-feynman} 
\tikzfeynmanset{compat=1.0.0}
\newcommand{\vtxdot}[1]{\fill (#1) circle (1.3pt);}
\usepackage{subcaption}

\usepackage{slashed}
\usepackage{graphicx}
\usepackage{bm}
\usepackage{amsmath}
\usepackage{accents}

\def\be{\begin{eqnarray}}
\def\ee{\end{eqnarray}}
\def\tr#1{\left\langle #1 \right\rangle}
\def\del{\partial}
\def\mrho{m_\rho}
\def\Mrho{M_\rho}

\def\zz#1{\accentset{\circ}{#1}}
\def\zm{\zz{m}_\pi}
\def\zf{\zz{f}_\pi}
\def\za{\zz{a}}
\def\zg{g}
\def\ar{a}
\def\aggff{\ar\, \zg^2 f_\pi^2}
\def\Az{A_\rho(0)}

\def\vmeson{double, double distance=0.5ex, line width=0.8pt}
\def\vmesonfull{double, double distance=1ex, line width=0.8pt}

\def\Tr#1{\left\langle{#1}\right\rangle}

\begin{document}
\title{Chiral perturbation theory with vector mesons
and the pion electromagnetic form factor}
\author{Tae-Sun Park}
\email{tspark@ibs.re.kr}
\affiliation{%
Center for Exotic Nuclear Studies, Institute for Basic Science, Daejeon 34126, Korea}

\date{\today}

\begin{abstract}
We formulate chiral perturbation theory with vector mesons kept as
explicit intermediate degrees of freedom, in the regime where the
momentum carried by a vector meson is of order $Q$ while its mass is of
order $\Lambda_\chi$.  This is the regime relevant to vector-meson
exchange in nuclear forces and low-energy nuclear electroweak processes.

We use the corresponding power counting and show that it gives a
consistent one-loop expansion, with the divergences generated by the
leading-order Lagrangian absorbed by next-to-leading-order
counterterms.

As an application, we calculate the pion electromagnetic form factor.
The vector mesons introduce a single new leading-order parameter,
$\ar$.  For the KSRF value $\ar=2$, the nonlocal one-loop contribution
to the direct photon--pion vertex vanishes at NLO, while the remaining
local term is found to be consistent with zero when fixed from the pion
charge radius and $\rho$-meson decay data.  Vector-meson dominance
therefore emerges as a good numerical approximation at this order.  The
explicit $\rho$ propagator, kept in its resummed form, also reproduces
the observed curvature of the form factor more efficiently than
vector-meson-less one-loop ChPT.  We also present a phenomenological extension of the form factor into
the $\rho$ resonance region, using a resummation of the $\rho$
self-energy.  The relation to other
formulations with explicit vector mesons is also discussed.
\end{abstract}


\maketitle

\section{Introduction}

Chiral perturbation theory (ChPT)
~\cite{wein68,ccwz1,ccwz,GL84,GL85,weinberg-counting,weinberg-counting2}
is the low-energy effective field theory of QCD based on chiral
symmetry.  In its usual form, heavy hadronic degrees of freedom are
integrated out and their effects are represented by local operators.
The expansion is organized in powers of $Q/\Lambda_\chi$, where $Q$
denotes a typical external momentum or the pion mass and
$\Lambda_\chi\simeq 4\pi f_\pi\simeq 1$ GeV.

At momenta well below the vector-meson mass, vector mesons can of course
be integrated out, with their effects encoded in local operators of ChPT.
When vector-meson exchange accounts for an important part of the dynamics,
however, this may not be the most efficient organization: the same physics
is then distributed over a sequence of higher-order operators.  Keeping the
vector mesons explicit provides a way to retain this structure while
preserving the low-energy chiral expansion.  A familiar example of the
general usefulness of keeping a heavier state explicit is the
$\Delta(1232)$ in nuclear ChPT~\cite{epj32,prc102,prc104}.

Several formulations of chiral theory with explicit vector mesons
have been developed.  Tensor-field formulations of massive vector
mesons were studied in Refs.~\cite{Ecker_NPB,Eckerl}, while the vector
realization of chiral symmetry was discussed by Georgi~\cite{georgi1990}.
In hidden local symmetry (HLS), the $\rho$ is introduced as a gauge
boson of a hidden symmetry and is counted as a light degree of freedom,
$m_\rho={\cal O}(Q)$~\cite{bando,bando-review,hls}.  In that
formulation, $\rho$ dominance of the pion form factor is known to
survive one-loop corrections for $\ar=2$~\cite{HaradaYamawaki1992}.
Heavy-vector-meson ChPT was developed in Refs.~\cite{jmw,bgt}, and
resonance chiral theory has also been applied to the pion vector form
factor at one loop using large-$N_C$ counting~\cite{rxt-loop}.  The
complex-mass scheme has been used for kinematics near the vector-meson
pole~\cite{cms,gegelia}, and other counting schemes with dynamical
vector mesons were considered in Refs.~\cite{lutz-leupold,terschluesen}.

Here we consider a different and more restricted kinematic regime.
The vector mesons occur only as intermediate states, and the momenta
they carry are small compared with their masses.  We use the counting
introduced in Refs.~\cite{tsp-report,mrho-counting}, in which the
momentum carried by a vector meson is of ${\cal O}(Q)$, while its mass
is of ${\cal O}(\Lambda_\chi)$.  This is the regime relevant, for
example, to vector-meson exchange in nuclear forces and nuclear
electroweak currents.  In this counting the vector-meson kinetic term
starts at next-to-leading order (NLO).  At leading order (LO) the
propagator is momentum independent and there is no $\rho\rho\rho$
vertex.  These features greatly simplify the loop expansion.  In
particular, the momentum dependence associated with vector-meson
propagation enters only at NLO, so that introducing explicit vector
mesons does not spoil the low-energy organization of the EFT.  The
loop organization is therefore different from that of HLS, in which
the vector-meson kinetic term is already present at LO.

The main purpose of this paper is to develop a consistent chiral
expansion with explicit vector mesons in this regime, including its
one-loop renormalization, and to apply it to the pion electromagnetic
form factor.  We identify the NLO counterterms required to absorb the
one-loop divergences and show explicitly that the expansion can be
renormalized at this order.

As an application, we calculate the pion electromagnetic form factor,
which is known through two loops in ordinary ChPT
~\cite{BCT1998,BT2002}.  The explicit vector-meson sector introduces
one new LO parameter, $\ar$; see Eq.~(\ref{calL0}) and the discussion
below it.  For the KSRF value $\ar=2$, the nonlocal one-loop contribution
to the direct photon--pion vertex vanishes at NLO.  The remaining local
term is fixed by the pion charge radius and the $\rho$-meson decay data.
A nontrivial result is that this term is found to be consistent with zero
and much smaller than its natural size.  Thus, for $\ar=2$, complete
vector-meson dominance (VMD) emerges as a good numerical approximation
at this order~\cite{gell-mann,sakurai}, rather than being imposed as an
assumption.  This point is discussed in detail in Sec.~\ref{sec-ee}.

We also compare the result with one-loop ChPT in which the vector
mesons have been integrated out.  No form-factor data at nonzero
momentum transfer are fitted.  The explicit $\rho$ propagator accounts
for the observed curvature of the form factor more efficiently than
the vector-meson-less one-loop result.  This comparison should not be
interpreted as a demonstration of faster order-by-order convergence,
since only one nontrivial order is calculated here.

For completeness, we also continue the form factor into the $\rho$
resonance region using a resummation of the one-loop $\rho$ self-energy.
This region lies outside the strict low-momentum counting, but the
resulting phenomenological extension is constrained to reproduce the
low-energy EFT through the order considered here.

The paper is organized as follows.  Section~2 presents the counting,
the Lagrangian, and the one-loop renormalization.  In
Sec.~\ref{sec-ee} we apply the formalism to the pion electromagnetic
form factor, compare it with the vector-meson-less result, and discuss
the resonance region.  Section~\ref{sec-summary} gives a summary.

\section{Formalism}

\subsection{Power counting rule and the Lagrangian}

We consider low-energy pion processes in which vector mesons occur only as
intermediate states.  We count a momentum carried by a vector meson as
$p={\cal O}(Q)$ and a vector-meson mass as $m_V={\cal O}(\Lambda_\chi)$.
The vector-meson propagator is then of order $1/\Lambda_\chi^2$.
The corresponding counting rule was derived in
Refs.~\cite{tsp-report,mrho-counting}.  A diagram with $E_H$ external
vector mesons, $E_e$ external gauge fields, and $L$ loops is of order
$Q^\nu$ (apart from powers of $f_\pi$ and $\Lambda_\chi$), with
\be
\nu = 2 - E_H - E_e + 2 L +
\sum_i \nu_i,
\qquad
\nu_i \equiv d_i + h_i + e_i -2,
\ee
where $d_i$, $h_i$, and $e_i$ are the numbers of derivatives and/or
pion-mass insertions, vector mesons, and external gauge fields at the
$i$-th vertex, respectively.

We expand the effective Lagrangian as
\be
{\cal L}= {\cal L}_0 + {\cal L}_2 + {\cal L}_4 + \cdots,
\ee
where the subscript on ${\cal L}$ denotes the chiral index $\nu_i$
of the vertices it contains.  We first recall
the LO pion Lagrangian of ordinary ChPT,
\be
{\cal L}_0^\text{ChPT} = \zf^2 \tr{i\Delta_\mu \, i \Delta^\mu}
+ \frac14 \zf^2 \tr{\chi_+},
\label{calL0pipi}\ee
with
\be
i\Gamma_\mu + i\Delta_\mu &=& i\xi^\dagger \del_\mu
\xi +\xi^\dagger ({\cal V}_\mu + {\cal A}_\mu) \xi,
\nonumber\\
i\Gamma_\mu - i\Delta_\mu &=& i\xi \del_\mu
\xi^\dagger + \xi ({\cal V}_\mu - {\cal A}_\mu) \xi^\dagger,
\nonumber\\
\chi_\pm &=& \xi^\dagger \chi  \xi^\dagger \pm
\xi \chi^\dagger \xi,
\ee
where 
${\cal V}_\mu$ and ${\cal A}_\mu$ are external vector and axial-vector
fields, while
$\chi= 2 B_0 ({\cal S}+ i {\cal P})$
contains the external scalar and pseudoscalar sources ${\cal S}$ and
${\cal P}$; $B_0$ is a constant,
and
$\tr{X}$ stands for the trace of $X$ in flavor space,
of dimension $N_f$,
$N_f$ being the number of flavors.  The scalar source ${\cal S}$
reduces in the vacuum to the quark-mass matrix.  In the explicit
pion--rho calculation below we work in the isospin-symmetric
$N_f=2$ sector, $m_u=m_d=\hat m$, and write
$\rho_\mu=\tau^a\rho_\mu^a/2$.  In the vacuum,
$\langle\chi\rangle=\zm^2 {\bf 1}$ with
$\zm^2=2B_0\hat m$.  The other external fields have vanishing
vacuum expectation values.
The parameters of the LO Lagrangian are denoted by the
``$\circ$" marks when they have to be distinguished from the
corresponding physical quantities.  See Ref.~\cite{GL85} for the
standard ChPT notation.
Here,
$\xi(x)$ is the square root of $U(x)$,
$U(x)$ being the $SU(N_f)$ field that contains the Goldstone bosons.
We use the parametrization
\be
U(x)=\exp\left(i \sum_a \lambda^a \pi^a(x)/f_\pi\right).
\ee
Here $f_\pi$ fixes the normalization of the pion coordinate in this
parametrization.  The coefficient $\zf$ of the LO chiral Lagrangian is
kept independent.  Its relation to the physical decay constant
$f_\pi$ is determined below after the NLO counterterms and loop
corrections are included.  With this convention the ratio
$\zf^2/f_\pi^2$ appears explicitly in the pion two-point function.
The generators are normalized by
$\tr{\lambda^a\lambda^b}=2\delta^{ab}$.
Under the chiral $\text{SU}(N_f)_L\times \text{SU}(N_f)_R$ transformation, 
$U$ transforms 
as 
$U\to g_R U g_L^\dagger$, 
and $g_{L,R}\in \text{SU}(N_f)_{L,R}$.
The non-linear realization 
of chiral symmetry~\cite{wein68,ccwz}
is obtained by the transformation properties of
the  $\xi(x)$:
\be
\xi\to g_R \,\xi\, h^\dagger = h \,\xi \,g_L^\dagger,
\ee
which defines $h= h[\xi, g_L, g_R]$.
The other quantities transform as
\be
\chi_\pm &\to& h \,\chi_\pm\, h^\dagger,
\nonumber\\
i\Delta_\mu &\to& h \,i \Delta_\mu\, h^\dagger,
\nonumber\\
i\Gamma_\mu &\to&
h \,i \Gamma_\mu\, h^\dagger + i h\, \del_\mu h^\dagger,
\ee
provided that the external fields transform as
${\cal V}_\mu + {\cal A}_\mu
\to g_{R} ({\cal V}_\mu + {\cal A}_\mu) g_{R}^\dagger
+ i g_{R} \, \del_\mu g_{R}^\dagger$,
${\cal V}_\mu - {\cal A}_\mu
\to g_{L} ({\cal V}_\mu - {\cal A}_\mu) g_{L}^\dagger
+ i g_{L} \, \del_\mu g_{L}^\dagger$
and $\chi\to g_R \,\chi\, g_L^\dagger$.


We include the massive vector mesons by extending the LO Lagrangian to
\be
{\cal L}_0=
{\cal L}_0^\text{ChPT} 
+ \za \zf^2
\tr{\left(\zg V_\mu - i \Gamma_\mu\right)^2}
+
\frac12 m_{\omega_1}^2 \omega_1^\mu \omega_{1\mu},
\label{calL0}\ee
where the $SU(3)$ notation is displayed for completeness; the
one-loop pion form-factor calculation below uses its $N_f=2$
$\rho$-meson sector.  In the three-flavor embedding,
\be
V_\mu= \sum_{a=1}^8 \frac{\lambda^a}{2} V^a_\mu
=
\frac{1}{\sqrt{2}}
\begin{pmatrix}
\frac{1}{\sqrt{2}} \rho^0_\mu+ \frac{1}{\sqrt{6}} \omega_{8\mu} & \rho^+_\mu & K^{*+}_\mu
\\
\rho^-_\mu & -\frac{1}{\sqrt{2}} \rho^0_\mu+ \frac{1}{\sqrt{6}} \omega_{8\mu} & K^{*0}_\mu \\
K^{*-}_\mu & \overline{K^{*0}}_\mu
& -\frac{2}{\sqrt{6}} \omega_{8\mu}
\end{pmatrix}
\ee
is the octet, and
$\omega_{1\mu}$ is the singlet 
vector field.
%
In the above equation, 
$\zg$ is the 
coupling constant
of the octet vector fields to the Goldstone bosons,
and $\za$ is dimensionless.  The coefficient of the quadratic vector
term contains the LO mass-scale parameter
$\za\zg^2\zf^2$.  Since there is no vector kinetic term at LO, this
quantity should not yet be identified with the physical pole mass.
After the NLO kinetic term and self-energy are included, the pole is
determined by Eq.~(\ref{rho-pole}).  The octet and singlet mesons are degenerate
in the large-$N_C$ limit~\cite{witten1979baryons},
$m_{\omega_1}^2=m_V^2$.
We assume 
the ideal $\phi-\omega$ mixing,
$|\phi\rangle = \cos\theta_V | \omega_8\rangle
- \sin\theta_V |\omega_1\rangle$
and
$|\omega\rangle = \sin\theta_V | \omega_8\rangle
+ \cos\theta_V |\omega_1\rangle$
with $\tan\theta_V=1/\sqrt{2}$,
which is implied by the Zweig rule~\cite{zweig}.

The vector fields transform as
\be
\zg V_\mu \to h \, \zg V_\mu\, h^\dagger + i h\, \del_\mu h^\dagger,
\ee
while the singlet field is unchanged, $\omega_1^\mu\to \omega_1^\mu$.
In this counting the usual vector-meson kinetic term is not part of
the LO Lagrangian, 
\be
{\cal L}_V^{\text{kin}} = -\frac12  \tr{V^{\mu\nu} V_{\mu\nu}} 
- \frac12 \omega_1^{\mu\nu} \omega_{1\mu\nu}
\ee
with $V_{\mu\nu}= \del_\mu V_\nu - \del_\nu V_\mu
- i g [V_\mu,\,V_\nu]$
and $\omega_{1\mu\nu}= \del_\mu \omega_{1\nu} - \del_\nu \omega_{1\mu}$,
%
It has $\nu_i=2$ and therefore first appears at NLO.  Without this
term, the LO vector-meson propagator is momentum independent,
$g^{\mu\nu}\delta^{ab}/m_V^2$.  The vector mesons can therefore be
integrated out algebraically at LO, giving ordinary ChPT.  At NLO the
kinetic term and loop corrections generate nontrivial momentum dependence.

Other countings are appropriate when vector mesons are external particles or when the momentum transfer is comparable to $m_V$.  Djukanovic et al. developed a scheme~\cite{cms,gegelia}
where the possible flows of the external momenta
through the internal propagators are considered,
assigning
the chiral order of a Feynman diagram
as the lowest
among all the possible cases.
In this scheme and in SU(2), the rho-meson kinetic term enters in the leading order
and the leading order rho-meson propagator is the familiar massive vector-meson propagator in the unitary gauge,
\be
\frac{1}{k^2 - \mrho^2 + i0^+} \left( - g_{\mu\nu} + \frac{k_\mu k_\nu}{\mrho^2}\right) \delta^{ab}.
\ee
However, this propagator
behaves poorly in the
ultraviolet region.
For example,
the diagram given in Fig.~\ref{Fig_rho3} with this propagator
results in a highly divergent term
of the form
$\frac{1}{d-4} k^4 (k^\mu k^\nu - g^{\mu\nu} k^2)$,
where $d$ is the space-time dimension.
\begin{figure}
\centering
   \begin{tikzpicture}[line width=0.8pt] \begin{feynman}
      \vertex (a); 
      \vertex [right=1cm of a] (c); \vertex [right=of c] (d);
      \vertex [right=1cm of d] (b);
      \diagram*{ 
      (a) --[\vmeson] (c);
      (d) --[\vmeson] (b);
      (c) -- [\vmeson, half left] (d) -- [\vmeson, half left] (c);
   };
   \end{feynman} \end{tikzpicture}
\caption{The most divergent one-loop diagram for the $\rho$ self-energy.
Double lines denote $\rho$ propagators.
\label{Fig_rho3}}
\end{figure}

In relativistic formulations, power-counting-violating analytic
pieces generated by heavy propagators can be separated from the
low-energy part and absorbed into LECs, for example by infrared
regularization~\cite{infra,bruns2005} or by the extended-on-mass-shell
scheme~\cite{extended}.  Both prescriptions have also been used in
calculations with explicit vector mesons~\cite{schindler2005}.
Another possibility is to
adopt the hidden local gauge symmetry (HLS), where
the rho meson is introduced as a gauge boson of the {\em hidden} symmetry~\cite{bando,hls}.
In HLS, one can adopt a general $R_\xi$ gauge propagator,
which removes the aforementioned problem.
HLS requires gauge fixing, which makes one-loop calculations more
involved.  A detailed one-loop analysis in the chiral limit is given in
Ref.~\cite{hls}.

The formalism presented in this paper
is free from the above-mentioned divergence,
since there is no $\rho\rho\rho$-vertex
at LO
and thus the aforementioned problematic one-loop diagram does not appear.

\subsection{One-loop divergences}

Before proceeding, we fix a notational convention.  The parameter of
the LO Lagrangian, Eq.~(\ref{calL0}), is written $\za$, while $\ar$
denotes the renormalized parameter whose relation to $\za$ is given in
Eq.~(\ref{ar-def}) below.
Since
$
\za-\ar={\cal O}\left(Q^2/\Lambda_\chi^2\right)$,
replacing $\za$ by $\ar$ inside a one-loop or NLO counterterm
contribution changes the result only at the next order.  We therefore
make this replacement throughout the NLO expressions that follow.
The replacement is not made in the LO tree-level vertices, where the
shift from $\za$ to $\ar$ is itself of the order being calculated and
is generated by the loop and counterterm contributions.
The same convention is used for $\zf$ and $\zm$: in loop and NLO
counterterm contributions we may replace $\zf$ by $f_\pi$ and
$\zm^2$ by $m_\pi^2$, while the circled quantities are retained in
LO tree-level terms when their NLO shifts are being determined.

The one-loop vacuum functional of pion ChPT was evaluated in
Refs.~\cite{GL84,GL85}.  Applying the same method with explicit vector
mesons gives the counterterm Lagrangian needed to absorb the divergences
generated by the LO Lagrangian,
\be
{\cal L}_{ct}
&=& \frac{1}{16\pi^2 (d-4)} \Bigg\{
 \frac{N_f}{6} \Tr{\tilde \Gamma_{\mu\nu}
 \tilde \Gamma^{\mu\nu}}
\nonumber\\
&+& 2 \Tr{\Delta^\mu\Delta^\nu}
\Tr{\Delta_\mu\Delta_\nu}
- \ar^2 \Tr{\Delta^\mu\phi^\nu}
\Tr{\Delta_\mu\phi_\nu}
\nonumber\\
&&
+ \frac{\ar^4}{8} \Tr{\phi^\mu\phi^\nu}
\Tr{\phi_\mu\phi_\nu}
\nonumber\\
&+& \Tr{W}^2
+ N_f \Tr{W^2}
- \frac{N_f}{2} \Tr{\chi_+ W} 
- \frac12 \Tr{\chi_+} \Tr{W}
\nonumber\\
&+& 
\frac{N_f^2-4}{16 N_f} \Tr{\chi_+^2}
+ \frac{N_f^2+2}{16 N_f^2} \Tr{\chi_+}^2
\Bigg\}
\nonumber \\
&+& \cdots,
\label{Lct}\ee
where the ellipses denote terms that are not needed for the vertex
functions considered below, 
$
\phi_\mu \equiv 
\left(\zg \rho_\mu - i \Gamma_\mu\right)$,
$W\equiv \Delta^\mu\Delta_\mu - \frac{\ar^2}{4} \phi^\mu\phi_\mu$,
and
\be
\tilde \Gamma_{\mu\nu} &=&
\left(1-\frac{\ar}{2}\right) \Gamma_{\mu\nu}
- i \frac{\ar}{2} \zg \rho_{\mu\nu}
\nonumber\\ &&
+ \frac{\ar}{2} \left(1-\frac{\ar}{2}\right)
\left[\zg \rho_\mu-i\Gamma_\mu,\, \zg \rho_\nu-i\Gamma_\nu\right]
\ee
with $\Gamma_{\mu\nu}
\equiv \partial_\mu \Gamma_\nu
- \partial_\nu \Gamma_\mu 
+ \left[\Gamma_\mu,\, \Gamma_\nu\right]$
and
$ \rho_{\mu\nu}
\equiv \partial_\mu \rho_\nu
- \partial_\nu \rho_\mu 
- i \zg \left[\rho_\mu,\, \rho_\nu\right]$.

\subsection{Renormalization of vertex functions}
We now renormalize the vertex functions needed for the pion electromagnetic form factor.
The relevant Lagrangian
may be written as
\be
{\cal L}_{2}
&=&
- \frac12 C_{1:1} \zg^2 \Tr{\rho_{\mu\nu}\rho^{\mu\nu}}
+\frac12 C_{1:2} \Tr{\Gamma_{\mu\nu}\Gamma^{\mu\nu}}
\nonumber\\ &&
-i C_{1:3} \zg \Tr{\rho^{\mu\nu}\Gamma_{\mu\nu}}
\nonumber\\
&& 
- i C_{2:1} \zg
\Tr{ \rho_{\mu\nu}\left[\Delta^\mu,\Delta^\nu\right]}
+C_{2:2} 
\Tr{\Gamma_{\mu\nu} \left[\Delta^\mu,\Delta^\nu\right]}
\nonumber\\
&&-  C_{3} \Tr{\chi_+}\Tr{\Delta^\mu\Delta_\mu}
+C_{4} \Tr{\chi_+} 
\Tr{(\zg \rho_\mu - i \Gamma_\mu)^2 }
\nonumber\\ &&
+ C_5 \Tr{\chi_+}^2 + \cdots,
\label{L2}\ee
where the ellipses represent terms that contain four or more
fields.
We use $C_i$ for the bare NLO low-energy constants (LECs).  In
dimensional regularization they are written as
\be
C_i=C_i^r(\mu)+\eta_i\lambda,
\label{Ci-ren}
\ee
where $\eta_i$ denotes the coefficient of the ultraviolet divergence
of the corresponding NLO operator, and
\be
\lambda \equiv
- \frac{\mu^{d-4}}{32\pi^2}
\left\{\frac{2}{4-d}+\ln4\pi+\Gamma'(1)+1\right\}.
\ee
For $\eta_i\ne0$, it is convenient to introduce the scale-independent
quantity $\bar C_i$ by
\be
C_i^r(\mu)=\frac{\eta_i}{32\pi^2}
\left(\bar C_i+\ln\frac{m_\pi^2}{\mu^2}\right).
\label{Cibar}
\ee
For $\eta_i=0$, however, Eq.~(\ref{Cibar}) is not used: the
corresponding $C_i^r$ is simply a finite LEC at this order.  This point
is relevant below for $C_{2:1}$ and $C_{2:2}$.

For the $N_f=2$ calculation, comparison of Eq.~(\ref{L2}) with
Eq.~(\ref{Lct}) gives
$\left\{\eta_{1:1},\eta_{1:2},\eta_{1:3}\right\}
=\frac23\left\{\frac{\ar^2}{4},
\left(1-\frac{\ar}{2}\right)^2,
\frac{\ar}{2}\left(1-\frac{\ar}{2}\right)\right\}$,
$\eta_{2:1}=\eta_{2:2}=0$,
$\left\{\eta_3,\eta_4\right\}
=\left\{1,\frac{\ar^2}{4}\right\}$, and
$\eta_5=3/32$.

The vector field requires a separate comment.  At LO only the
combination $g\rho_\mu$ occurs and there is no canonical kinetic term,
so the separate normalization of $g$ and $\rho_\mu$ is not fixed at
that order.  At the order considered here no separate bare coupling
$g_0$ or vector-field renormalization constant is required.  The
ultraviolet divergence of the transverse vector two-point function is
absorbed by $C_{1:1}$.  We then choose the finite normalization of the
vector field through the pole-residue condition, which fixes the
finite value of $g$.

\subsubsection{1PI for ${\pi}_a(k) \to {\pi}_b(k)$}
\begin{figure}
\centering
\begin{subfigure}{0.2\textwidth} \centering
   \begin{tikzpicture}[line width=0.8pt] \begin{feynman}
      \vertex (a); \vertex [right=0.7cm of a] (c); \vertex [right=1.4cm of c] (d);
      \vertex [right=0.7cm of d] (b);
      \diagram*{
      (a) --[scalar] (c);
      (d) --[scalar] (b);
      (c) -- [\vmeson] (d);
      (c) -- [scalar, half left] (d);
   };
   \vtxdot{c}\vtxdot{d}
   \end{feynman} \end{tikzpicture}
   \caption{} \end{subfigure}
\begin{subfigure}{0.2\textwidth} \centering
   \begin{tikzpicture}[line width=0.8pt] \begin{feynman}
      \vertex (a); \vertex [right=1.2cm of a] (c); \vertex [right=1.2cm of c] (b);
      \vertex [above=1.2cm of c] (d);
      \diagram*{
      (a) -- [scalar] (c) --[scalar] (b);
      (c) -- [scalar, half left] (d) --[scalar, half left] (c);
   };
   \vtxdot{c}
   \end{feynman} \end{tikzpicture}
   \caption{} \end{subfigure}
\caption{1PI diagrams for ${\pi}_a(k) \to {\pi}_b(k)$.
The dashed lines denote pions.
\label{Fig_pipi}}
\end{figure}

The one-particle irreducible (1PI) vertex for $\pi_a(k) \to \pi_b(k)$
is the inverse of the pion propagator
$\delta_{ab}\, D_\pi(k^2)$,
whose one-loop diagrams
are drawn in Fig.~\ref{Fig_pipi}.
Their contributions are
\be
D_\pi^{-1, (a)}(k^2) &=& 2 \ar k^2 \frac{\lambda_2}{f_\pi^2},
\nonumber \\
D_\pi^{-1, (b)}(k^2) &=& 
\left[ 2 (1-\ar) k^2 - \frac{3}{2} m_\pi^2\right]
\frac{\lambda_2}{f_\pi^2},
\label{pipi-loop}\ee
where 
\be
\lambda_2 
&\equiv& 
-2 m_\pi^2 \left(
      \lambda + \frac{1}{32\pi^2} \ln\frac{m_\pi^2}{\mu^2}\right).
\ee
There are also {\em tree} contributions from ${\cal L}_0$ and ${\cal L}_2$,
\be
D_\pi^{-1, \text{tree}}(k^2) &=& 
\frac{\zf^2}{f_\pi^2}\left(k^2-\zm^2\right)
+\frac{4m_\pi^2 C_3}{f_\pi^2}\,k^2
-\frac{32m_\pi^4 C_5}{f_\pi^2}
\nonumber\\
&\equiv& (1+\delta Z_\pi) k^2 - (m_\pi^2+\delta m_\pi^2),
\label{pipi-tree}\ee
which defines $\delta Z_\pi$ and $\delta m_\pi^2$.
Adding the loop and tree contributions gives
\be
D_\pi^{-1}(k^2) &\equiv&
D_\pi^{-1, (a)}(k^2)+ D_\pi^{-1, (b)}(k^2)
+D_\pi^{-1, \text{tree}}(k^2)
\nonumber\\
&=& k^2 - m_\pi^2
\ee
with
\be
\delta Z_\pi &=& -2 \frac{\lambda_2}{f_\pi^2},
\nonumber\\
\delta m_\pi^2 &=& -\frac{3}{2} \frac{\lambda_2}{f_\pi^2} m_\pi^2\,.
\label{f0m0}\ee
Combining Eqs.~(\ref{pipi-tree}) and (\ref{f0m0}) gives
\be
f_\pi^2 &=&
\zf^2 + \frac{\zm^2}{8\pi^2} \bar C_3,
\nonumber\\
m_\pi^2 
&=& \zm^2 - \frac{\zm^4}{32\pi^2 f_\pi^2} (4 \bar C_3 - 3 \bar C_5).
\label{f0m1}\ee
Here we keep the LO quantities in the NLO corrections in
Eq.~(\ref{f0m1}) to display explicitly their relation to the physical
parameters.  They may equivalently be replaced by the corresponding
physical quantities at this order.
This agrees with Ref.~\cite{GL84}: in the present convention,
$C_3$ and $C_5$ correspond to $l_4/2$ and $(l_3+l_4)/16$,
respectively.

The 1PI functions for 
${\pi}_b(k) \to {\cal A}_a^\mu(k)$
and ${\cal A}_a^\mu(k) \to {\cal A}_b^\nu(k)$ can be evaluated in a similar way:
\be
\Gamma\left[
{\pi}_a(k) \to {\cal A}_b^\mu(k)\right] 
&=& i k^\mu f_\pi\, \delta_{ab},
\\
\Gamma\left[
{\cal A}_a^\mu(k) \to {\cal A}_b^\nu(k)\right] 
&=& f_\pi^2 \, g^{\mu\nu}\,\delta_{ab}.
\ee

\subsubsection{1PI for ${\cal V}_e^\mu(k) \to \pi_a(q_a) + \pi_b(q_b)$}
\begin{figure}
\centering
\begin{subfigure}{0.15\textwidth} \centering
   \begin{tikzpicture}[line width=0.8pt] \begin{feynman}
   \vertex (e); \vertex [right=0.7cm of e] (c); \vertex [right=0.9cm of c] (d);
   \vertex [below=1.2cm of c] (a); \vertex [above=1.2cm of c] (b);
   \diagram*{
      (e) --[photon] (c);
      (a) --[scalar] (c) --[scalar] (b);
      (c) --[scalar, half left] (d) --[scalar, half left] (c);
   }; 
   \vtxdot{c}
   \end{feynman} \end{tikzpicture}
   \caption{} \end{subfigure}
\begin{subfigure}{0.15\textwidth} \centering
   \begin{tikzpicture}[line width=0.8pt] \begin{feynman}
   \vertex (e); \vertex [right=0.7cm of e] (v); \vertex [right=0.9cm of v] (c);
   \vertex [below=1.2cm of c] (a); \vertex [above=1.2cm of c] (b);
   \diagram*{
      (e) --[photon] (v);
      (a) --[scalar] (c) --[scalar, half left] (v) --[scalar, half left] (c) -- [scalar] (b);
   };
   \vtxdot{v}\vtxdot{c}
   \end{feynman} \end{tikzpicture}
   \caption{} \end{subfigure}
\begin{subfigure}{0.15\textwidth} \centering
   \begin{tikzpicture}[line width=0.8pt] \begin{feynman}
   \vertex (e); \vertex [right=0.7cm of e] (v); \vertex [right=0.8cm of v] (h);
   \vertex [below=0.6cm of h] (c); \vertex [above=0.6cm of h] (d);
   \vertex [below=1.2cm of h] (a); \vertex [above=1.2cm of h] (b);
   \diagram*{
      (e) --[photon] (v);
      (a) --[scalar] (c) --[scalar, half left, looseness=0.5] (v) 
         -- [scalar, half left, looseness=0.5] (d) -- [scalar] (b);
      (c) --[\vmeson] (d);
   };
   \vtxdot{v}\vtxdot{c}\vtxdot{d}
   \end{feynman} \end{tikzpicture}
   \caption{} \end{subfigure}
\begin{subfigure}{0.15\textwidth} \centering
   \begin{tikzpicture}[line width=0.8pt] \begin{feynman}
   \vertex (e); \vertex [right=0.7cm of e] (c);
   \vertex [above=1.2cm of c] (d);
   \vertex [below=0.6cm of c] (a); \vertex [above=0.6cm of d] (b);
   \diagram*{
      (a) --[scalar] (c) --[scalar, half right] (d) -- [scalar] (b);
      (c) --[\vmeson] (d);
      (e) --[photon] (c);
   };
   \vtxdot{c}\vtxdot{d}
   \end{feynman} \end{tikzpicture}
   \caption{} \end{subfigure}
\begin{subfigure}{0.15\textwidth} \centering
   \begin{tikzpicture}[line width=0.8pt] \begin{feynman}
   \vertex (e); \vertex [right=0.7cm of e] (d);
   \vertex [below=1.2cm of d] (c);
   \vertex [below=0.6cm of c] (a); \vertex [above=0.6cm of d] (b);
   \diagram*{
      (a) --[scalar] (c) --[scalar, half right] (d) -- [scalar] (b);
      (c) --[\vmeson] (d);
      (e) --[photon] (d);
   };
   \vtxdot{c}\vtxdot{d}
   \end{feynman} \end{tikzpicture}
   \caption{} \end{subfigure}
\caption{1PI diagrams for ${\cal V}_e^\mu(k)\to \pi_a(q_a) + \pi_b(q_b)$.
The wavy lines denote external vector fields.
\label{fig-Vpipi}}
\end{figure}

The 1PI diagrams for
${\cal V}_e^\mu(k) \to\pi_a(q_a) + \pi_b(q_b)$
are drawn in Fig.~\ref{fig-Vpipi}.  Momentum conservation gives
$k=q_a+q_b$, and from here on we write $t\equiv k^2$.
The corresponding vertex function can be written as
$-i\epsilon_{eab}(q_a-q_b)^\mu\Gamma_{{\cal V}\pi\pi}(t)$,
where $\epsilon_{eab}$ is the totally antisymmetric SU(2) tensor.
The dimensionally regulated pion-loop function $f_3(t)$ is defined in
Eq.~(\ref{f3-def}) below.  The individual
contributions are
\be
\Gamma_{{\cal V}\pi\pi}^{(a)}(t)&=& 
3 (1-\ar) \frac{\lambda_2}{f_\pi^2},
\nonumber\\
\Gamma_{{\cal V}\pi\pi}^{(b)}(t)&=& 
-\frac{(2-\ar)^2}{4 f_\pi^2}
\left[
\lambda_2 
- t f_3(t) \right],
\nonumber\\
\Gamma_{{\cal V}\pi\pi}^{(c)}(t)&=&0,
\nonumber\\
\Gamma_{{\cal V}\pi\pi}^{(d+e)}(t)&=& 
2 \ar \frac{\lambda_2}{f_\pi^2},
\nonumber\\
\Gamma^{\text{tree}}_{{\cal V}\pi\pi}(t) &=&
1+\delta Z_\pi - 
   \frac{1}{2 f_\pi^2}\left(\za \zf^2 + 4 m_\pi^2 C_4\right)
\nonumber\\
&+&
 \frac{t}{2 f_\pi^2} \left(C_{1:2}-C_{2:2}\right),
\ee

Adding the diagrams gives
\be
\Gamma_{{\cal V}\pi\pi}(t)
&=&
1-\frac{\ar}{2}
\nonumber\\
&+&\frac{t}{2 f_\pi^2} \left[
2 \left(1-\frac{\ar}{2}\right)^2  f_3(t) 
+ C_{1:2}-C_{2:2}\right]
\label{Vpipi}\ee
Equation~(\ref{Vpipi}) is the convenient un-subtracted form: $f_3$ and
$C_{1:2}$ separately contain a divergence.  To make the
renormalization explicit.  Using
$f_3(t)=\bar f_3(t)+f_3(0)$ and Eq.~(\ref{Ci-ren}), the same result can
be written as
\be
\Gamma_{{\cal V}\pi\pi}(t)
&=&1-\frac{\ar}{2}
+\frac{t}{2f_\pi^2}\Bigg\{
2\left(1-\frac{\ar}{2}\right)^2\bar f_3(t)
+C_{1:2}^r(\mu)-C_{2:2}
\nonumber\\&&
-\frac{\left(1-\frac{\ar}{2}\right)^2}{48\pi^2}
\left(\ln\frac{m_\pi^2}{\mu^2}-1\right)\Bigg\}.
\label{Vpipi-finite}\ee
The expression in braces is finite and scale independent.  The term
proportional to $\bar f_3(t)$ in Eq.~(\ref{Vpipi-finite}) is the
nonlocal part: it contains the nonanalytic momentum dependence
generated by the pion loop, including the two-pion branch cut.  The
remaining terms are local in the external momentum at this order: they
multiply $t$ polynomially and are represented by local NLO operators
after renormalization.  If $\ar\ne2$, this local part may equivalently be
written as $\left(1-\ar/2\right)^2(\bar C_{1:2}+1)/(48\pi^2)-C_{2:2}$.
If $\ar=2$, however, $\eta_{1:2}=0$ and $C_{1:2}$ itself is a finite
LEC; no $\bar C_{1:2}$ is introduced.  The LO and renormalized parameters are related by the finite NLO shift
\be
\za &=& \ar+\delta a,
\nonumber\\
\delta a &=&
\frac{m_\pi^2}{8\pi^2 f_\pi^2}
\left(\ar \bar C_3 - \frac{\ar^2}{4}\bar C_4\right)
+{\cal O}(Q^4/\Lambda_\chi^4).
\label{ar-def}\ee
Here $\delta a$ is a finite NLO shift, not an additional independent
counterterm; the ultraviolet divergences have already been absorbed by
the NLO LECs $C_i$.
If $\ar=0$ the results reduce to ChPT without vector mesons.  The choice
$\ar=2$, corresponding to the KSRF relation~\cite{ksrf1,ksrf2},
simplifies them instead: the direct coupling at the origin,
$\Gamma_{{\cal V}\pi\pi}(0)=1-\ar/2$, then vanishes, and the nonlocal
one-loop term proportional to $(1-\ar/2)^2f_3(t)$ drops out at this
order as well.

The condition may be imposed on $\ar$ or on $\za$.  The two differ by
$\mathcal{O}(Q^2/\Lambda_\chi^2)$, and in the form factor the resulting
shift of the direct vertex is compensated by the shift of the
normalization of the pole term, Eq.~(\ref{Fpit}), so that $F_\pi(t)$
changes only at $\mathcal{O}(Q^4/\Lambda_\chi^4)$.  We impose $\ar=2$,
which makes $\Gamma_{{\cal V}\pi\pi}(0)$ vanish exactly.

A finite local NLO term $C_{1:2}-C_{2:2}$ remains.  Complete VMD is
therefore not imposed at this stage; the remaining finite combination
is determined in Sec.~\ref{sec-ee} from the pion charge radius together
with the $\rho$ decay data.  For general $\ar$ the divergence of the
loop term is cancelled by that of $C_{1:2}$, whereas $C_{2:2}$ is finite
at this order.

\subsubsection{1PI for ${\rho}_e^\mu(k)\to \pi_a(q_a) + \pi_b(q_b)$}
\begin{figure}
\centering
\begin{subfigure}{0.15\textwidth} \centering
   \begin{tikzpicture}[line width=0.8pt] \begin{feynman}
   \vertex (e); \vertex [right=0.7cm of e] (c); \vertex [right=0.9cm of c] (d);
   \vertex [below=1.2cm of c] (a); \vertex [above=1.2cm of c] (b);
   \diagram*{
      (e) --[\vmeson] (c);
      (a) --[scalar] (c) --[scalar] (b);
      (c) --[scalar, half left] (d) --[scalar, half left] (c);
   };
   \vtxdot{c}
   \end{feynman} \end{tikzpicture}
   \caption{} \end{subfigure}
%
\begin{subfigure}{0.15\textwidth} \centering
   \begin{tikzpicture}[line width=0.8pt] \begin{feynman}
   \vertex (e); \vertex [right=0.7cm of e] (v); \vertex [right=0.9cm of v] (c);
   \vertex [below=1.2cm of c] (a); \vertex [above=1.2cm of c] (b);
   \diagram*{
      (e) --[\vmeson] (v);
      (a) --[scalar] (c) --[scalar, half left] (v) --[scalar, half left] (c) -- [scalar] (b);
   };
   \vtxdot{v}\vtxdot{c}
   \end{feynman} \end{tikzpicture}
   \caption{} \end{subfigure}
%
\begin{subfigure}{0.15\textwidth} \centering
   \begin{tikzpicture}[line width=0.8pt] \begin{feynman}
   \vertex (e); \vertex [right=0.7cm of e] (v); \vertex [right=0.8cm of v] (h);
   \vertex [below=0.6cm of h] (c); \vertex [above=0.6cm of h] (d);
   \vertex [below=1.2cm of h] (a); \vertex [above=1.2cm of h] (b);
   \diagram*{
      (e) --[\vmeson] (v);
      (a) --[scalar] (c) --[scalar, half left, looseness=0.5] (v)
         -- [scalar, half left, looseness=0.5] (d) -- [scalar] (b);
      (c) --[\vmeson] (d);
   };
   \vtxdot{v}\vtxdot{c}\vtxdot{d}
   \end{feynman} \end{tikzpicture}
   \caption{} \end{subfigure}
%
\caption{1PI diagrams for ${\rho}_e^\mu(k) \to\pi_a(q_a) + \pi_b(q_b)$.
The double lines denote $\rho$ mesons.
\label{fig-rhopipi}}
\end{figure}
The 1PI diagrams
for ${\rho}_e^\mu(k)\to \pi_a(q_a) + \pi_b(q_b)$
are drawn in Fig.~\ref{fig-rhopipi},
whose vertex function
can be written as
$-i\epsilon_{eab} (q_a-q_b)^\mu \Gamma_{\rho\pi\pi}(k^2)$
with
\be
\Gamma_{\rho\pi\pi}^{(a)}&=& \frac{\ar \zg}{2 f_\pi^2}  \lambda_2,
\nonumber\\
\Gamma_{\rho\pi\pi}^{(b)}&=& 
\left(1-\frac{\ar}{2}\right)
\frac{\ar \zg }{2 f_\pi^2}
\left[-\lambda_2 + t f_3(t)\right],
\nonumber\\
\Gamma_{\rho\pi\pi}^{(c)}&=&0,
\nonumber\\
\Gamma_{\rho\pi\pi}^{\text{tree}}
&=&
\frac{\zg}{2 f_\pi^2}
\left\{
\za \zf^2 + 4 m_\pi^2 C_4
+ 
\left(
C_{1:3} - C_{2:1}\right) t
\right\},
\ee
and their sum is
\be
\Gamma_{\rho\pi\pi}(t)
=
\frac{\ar}{2} \zg 
+ \frac{\zg}{2 f_\pi^2} t \left[
\ar \left(1-\frac{\ar}{2}\right) f_3(t) 
+ C_{1:3} - C_{2:1} 
\right].
\ee
The explicitly finite form is
\be
\Gamma_{\rho\pi\pi}(t)
&=&\frac{\ar}{2}\zg
+\frac{\zg t}{2f_\pi^2}\Bigg\{
\ar\left(1-\frac{\ar}{2}\right)\bar f_3(t)
+C_{1:3}^r(\mu)-C_{2:1}
\nonumber\\&&
-\frac{\ar\left(1-\frac{\ar}{2}\right)}{96\pi^2}
\left(\ln\frac{m_\pi^2}{\mu^2}-1\right)\Bigg\},
\label{rhopipi-finite}\ee
For $\ar=2$, the LECs $C_{1:3}$ and $C_{2:1}$ are finite at this
order, and the nonlocal loop term does not contribute at NLO.  We then
obtain
\be
\left.\Gamma_{\rho\pi\pi}(t)\right|_{\ar=2}
&=&
\zg 
+ \frac{t}{m_\rho^2} \left( g_{\rho\pi\pi} - \zg\right),
\ee
where
\be
g_{\rho\pi\pi} \equiv \Gamma_{\rho\pi\pi}(m_\rho^2) = \zg \left[
 1 + \frac{m_\rho^2}{2 f_\pi^2}  \left(
C_{1:3} - C_{2:1} 
 \right)\right]
\label{grhopipi-form} \ee
is fixed below.  To determine numerical values of the finite
$\rho$-sector LECs we use observables at the $\rho$ pole.  Evaluating
the NLO vertex functions at $t=m_\rho^2$ is outside the strict
low-$Q$ expansion and should be regarded as a phenomenological
matching prescription.  Once the LECs are fixed, their use in the
low-momentum form factor is within the counting described above.
Since
$\Gamma(\rho^0 \to \pi^+\pi^-)= g_{\rho\pi\pi}^2 p_\pi^3/(6\pi m_\rho^2)$
with $p_\pi\equiv \sqrt{m_\rho^2/4-m_\pi^2}$,
approximating the neutral $\rho$ width by its dominant
$\pi^+\pi^-$ mode and using
$\Gamma_\rho=(147.4\pm0.8)$ MeV~\cite{pdg2026} gives
\be
g_{\rho\pi\pi}= 5.942\pm 0.017.
\label{grhopipi}\ee

\subsubsection{1PI for ${\rho}_a^\mu(k) \to {\rho}_b^\nu(k)$}

\begin{figure}[bt]
\centering
\begin{subfigure}{0.2\textwidth} \centering
   \begin{tikzpicture}[line width=0.8pt] \begin{feynman}
   \vertex (a); \vertex [right=0.7cm of a] (c); \vertex [right=1.2cm of c] (d);
   \vertex [right=0.7cm of d] (b);
   \diagram*{
      (a) --[\vmeson] (c);
      (d) --[\vmeson] (b);
      (c) -- [scalar, half left] (d) -- [scalar, half left] (c);
   };
   \vtxdot{c}\vtxdot{d}
   \end{feynman} \end{tikzpicture}
   \caption{} \end{subfigure}
\begin{subfigure}{0.2\textwidth} \centering
   \begin{tikzpicture}[line width=0.8pt] \begin{feynman}
   \vertex (a); \vertex [right=1.2cm of a] (c); \vertex [right=1.2cm of c] (b);
   \vertex [above=1.2cm of c] (d);
   \diagram*{
      (a) -- [\vmeson] (c) --[\vmeson] (b);
      (c) -- [scalar, half left] (d) --[scalar, half left] (c);
   };
   \vtxdot{c}
   \end{feynman} \end{tikzpicture}
   \caption{} \end{subfigure}
\caption{1PI diagrams for ${\rho}_a^\mu(k) \to {\rho}_b^\nu(k)$.
\label{Fig_rhorho}}
\end{figure}

The vertex function
for ${\rho}_a^\mu(k) \to {\rho}_b^\nu(k)$ 
is the inverse of the
rho-meson propagator $\delta_{ab} D_\rho^{\mu\nu}(k)$,
whose relevant diagrams 
are drawn in
Fig.~\ref{Fig_rhorho}.
The loop contributions read
\be
\left(D_{\rho}^{-1}\right)^{\mu\nu, (a)}(k)
&=&\frac12  \ar^2 \zg^2 \lambda_2 \, g^{\mu\nu}
\nonumber\\ &&
+ \frac12  \ar^2 \zg^2 \left(k^\mu k^\nu - g^{\mu\nu} k^2\right) f_3(k^2),
\nonumber\\
\left(D_{\rho}^{-1}\right)^{\mu\nu, (b)}(k) &=& 0,
\nonumber\\
\left(D_{\rho}^{-1}\right)^{\mu\nu, \text{tree}}(k) &=& 
\zg^2 \left(\za \zf^2 
+ 4 \zm^2 C_4\right) g^{\mu\nu} 
\nonumber\\ &&
+ 
\zg^2 C_{1:1} \left(k^\mu k^\nu - g^{\mu\nu}k^2\right),
\ee
The resulting propagator is
\be
D^{\mu\nu}_\rho(k) 
&=& - D_\rho(k^2) \left( g^{\mu\nu} - A_\rho(k^2) \frac{k^\mu k^\nu}{\aggff}
\right)
\label{rho-prop}\ee
with
\be
D_\rho(t) &=& 
\left[t A_\rho(t) - \aggff\right]^{-1},
\\
A_\rho(t)
&=&
\zg^2 C_{1:1}+
\frac12\ar^2\zg^2 f_3(t)
\nonumber\\
&=& A_\rho(0)+\frac12\ar^2\zg^2\bar f_3(t),
\label{Arho-ren}
\ee
where the second line is finite: the divergence in $f_3(0)$ is
absorbed by $C_{1:1}$ according to Eq.~(\ref{Ci-ren}).  For $N_f=2$
and $\eta_{1:1}=\ar^2/6$, this cancellation can also be written as
\be
A_\rho(0)=\zg^2\left[C_{1:1}^r(\mu)
-\frac{\ar^2}{192\pi^2}
\left(\ln\frac{m_\pi^2}{\mu^2}-1\right)\right]
=\frac{\ar^2\zg^2}{192\pi^2}(\bar C_{1:1}+1),
\label{Arho0-finite}\ee
where the last form applies when $\ar\ne0$.

We fix $g$ and $C_{1:1}$ by requiring the propagator to have the following form near the pole,
\be
D^{-1}_\rho(t)
= (t-\mrho^2)+i m_\rho\Gamma_\rho
+{\cal O}\!\left((t-\mrho^2)^2\right),
\label{rho-pole}\ee
where $\Gamma_\rho \simeq \Gamma(\rho^0 \to \pi^+\pi^-)$
is the decay width of the rho meson.
%
For the phenomenological continuation to the resonance region, we
replace
$\bar A_\rho(t) \equiv A_\rho(t)-\Az$ by
\be
\bar A_\rho(t) 
= 2 \left[\Gamma_{\rho\pi\pi}(t)\right]^2
\bar f_3(t).
\label{Arhot}\ee
Here the tree-level coupling $\ar g/2$ in the one-loop self-energy is
replaced by the full NLO vertex $\Gamma_{\rho\pi\pi}(t)$.  Since
$\Gamma_{\rho\pi\pi}(t)=\ar g/2+{\cal O}(Q^2/\Lambda_\chi^2)$ for
$t={\cal O}(Q^2)$, this replacement changes the strict low-energy
result only beyond NLO.  Near the pole it supplies the higher-order
terms needed to reproduce the physical $\rho$ width.
For the numerical analysis we now set $\ar=2$.  We define
$R_3\equiv \Re\bar f_3(m_\rho^2)$,
$R_3'\equiv \Re f_3'(m_\rho^2)$, and
$f_3'(t)\equiv d f_3(t)/dt$.
The conditions
$\aggff=t\,\Re A_\rho(t)$ and
$\partial[t\,\Re A_\rho(t)]/\partial t=1$ at $t=\mrho^2$
then give, within this phenomenological replacement and after dropping
explicit terms quadratic in the one-loop function,\footnote{\protect
If one does not invoke the replacement given in Eq.~(\ref{Arhot}),
the corresponding solution at $\ar=2$ is
\be
\zg &=& \left( \frac{2 f_\pi^2}{m_\rho^2}
+ 2 m_\rho^2 R_3'\right)^{-1/2}
\simeq 6.23\pm 0.01,
\nonumber \\
\Az&=& \frac{2 \zg^2 f_\pi^2}{m_\rho^2}
-2 \zg^2 R_3
\simeq 1.126\pm 0.001.
\ee
}
\be
\zg &=& \frac{m_\rho}{\sqrt{2} f_\pi}
\Big[
\frac{\sqrt{2}g_{\rho\pi\pi} m_\rho}{f_\pi} R_3
\nonumber\\&&
+
\sqrt{1 - 2 g_{\rho\pi\pi}^2 \left(
 2 R_3  + m_\rho^2 R_3'
\right)}
\Big]
\nonumber\\
&\simeq& 6.21\pm 0.06,
\nonumber\\
\Az &=& \frac{2 \zg^2 f_\pi^2}{m_\rho^2} - 2 g_{\rho\pi\pi}^2 R_3
\simeq 1.112\pm 0.001,
\label{g}\ee
where numerically
$R_3\simeq -0.3395\times10^{-3}$ and
$R_3'\simeq -1.289\times10^{-3}/m_\rho^2$.
We use $m_\rho=(775.26\pm0.23)$ MeV and
$f_\pi = f_{\pi^+}/\sqrt2 = (92.07\pm0.85)$ MeV, from
$f_{\pi^+}=(130.2\pm1.2)$ MeV~\cite{pdg2026}.  All uncertainties quoted
below are obtained by propagating the errors of $m_\rho$, $f_\pi$, the
$\rho$ widths and the pion charge radius through the expressions given
here.

For small $k^2$, neglecting terms that first affect the propagator at
higher order in $k^2$, one may write
\be
D^{\mu\nu}_\rho(k)
= \frac{1}{\Az}
\frac{1}{\tilde \Mrho^2-k^2}
\left(g^{\mu\nu}-\frac{k^\mu k^\nu}{\tilde\Mrho^2}\right)
+{\cal O}(k^4/\Lambda_\chi^4),
\ee
where
\be
\tilde \Mrho \equiv \frac{\sqrt{\ar} \zg f_\pi}{\sqrt{\Az}}\
\simeq (0.989 \pm 0.001) \ \mrho.
\ee
Thus the overall normalization is reduced by
$1/\Az\simeq0.90$, and the effective mass is about one percent below
the physical mass, $\tilde\Mrho\simeq0.99\mrho$.

The 1PI $\rho$--${\cal V}$ mixing amplitude,
${\rho}_a^\mu(k) \to {\cal V}_b^\nu(k)
\equiv \Gamma_{{\rho}{\cal V}}^{\mu\nu}(k)$, is
\be
\Gamma_{{\rho}{\cal V}}^{\mu\nu}(k)
&=&
- \Gamma_{\rho\cal{V}}(k^2) g^{\mu\nu} 
\nonumber\\ 
&+& \zg \left[\ar \left(1- \frac{\ar}{2}\right)
f_3(k^2) + C_{1:3} \right] k^\mu k^\nu
\ee
with
\be
\Gamma_{\rho\cal{V}}(t) = \ar \zg f_\pi^2 
+ \zg \left[\ar \left(1- \frac{\ar}{2}\right) 
f_3(t) + C_{1:3} \right] t .
\label{rhoVt}\ee
Equivalently, its transverse scalar coefficient can be written in the
finite form
\be
\Gamma_{\rho\cal{V}}(t)
&=&\ar\zg f_\pi^2+\zg t\Bigg\{
\ar\left(1-\frac{\ar}{2}\right)\bar f_3(t)+C_{1:3}^r(\mu)
\nonumber\\&&
-\frac{\ar\left(1-\frac{\ar}{2}\right)}{96\pi^2}
\left(\ln\frac{m_\pi^2}{\mu^2}-1\right)\Bigg\},
\label{rhoV-finite}\ee
For $\ar=2$, $C_{1:3}$ is finite and the loop term starts beyond
NLO.
$\Gamma_{\rho\cal{V}}(m_\rho^2)$ is related to the $\rho\to e^+e^-$ decay
width as
\be
\Gamma(\rho\to e^+e^-)
=\frac{8\pi\alpha^2}{3m_\rho^4}
\left|\Gamma_{\rho\cal V}(m_\rho^2)\right|^2
\left(1+2\frac{m_e^2}{m_\rho^2}\right)p_e.
\ee
Here $\alpha$ is the fine-structure constant, $m_e$ is the electron
mass, and $p_e\equiv \sqrt{m_\rho^2/4-m_e^2}$ is the electron momentum
in the $\rho$ rest frame.
The PDG partial width is
$\Gamma(\rho^0\to e^+e^-)=(7.04\pm0.06)$ keV~\cite{pdg2026}, which
gives
\be
\Gamma_{\rho\cal{V}}(m_\rho^2)=(0.1213\pm0.0005)\ \text{GeV}^2.
\label{rhoVvalue}\ee
Numerically,
$\Gamma_{\rho\cal{V}}(m_\rho^2)/(2f_\pi^2g_{\rho\pi\pi})\simeq1.20$.
For $\ar=2$, Eqs.~(\ref{grhopipi-form},\ref{grhopipi},
\ref{rhoVt},\ref{rhoVvalue}) first determine the two finite
combinations
\be
C_{1:3} &=& (4.27 \pm 0.27)\times 10^{-3},
\nonumber\\
C_{1:3}-C_{2:1}
&=& (-1.22 \pm 0.23)\times 10^{-3}.
\label{C13comb}\ee
Combining these results, with the correlations from the common inputs
retained, gives
\be
C_{2:1}=(5.49\pm0.48)\times10^{-3}.
\label{C21}\ee

The correlations are important here.  The quantities in
Eq.~(\ref{C13comb}) share common inputs through $\zg$, which is itself
a derived quantity determined by $f_\pi$, $g_{\rho\pi\pi}$ and
$m_\rho$ through Eq.~(\ref{g}).  Consequently, the derived
$C_{1:3}$ and $C_{2:1}$ are strongly correlated; their correlation
coefficient is $0.96$.  This strong correlation is why the combination
$C_{1:3}-C_{2:1}$ is much better determined than one would infer by
treating the individual errors as independent.

The quoted uncertainties are obtained by taking as independent the
five measured inputs
\be
m_\rho,\quad f_\pi,\quad \Gamma_\rho,\quad
\Gamma(\rho^0\to e^+e^-),\quad \langle r_\pi^2\rangle ,
\ee
sampling them from independent Gaussians with the uncertainties quoted
above, and evaluating the full chain
$g_{\rho\pi\pi} \to \Gamma_{\rho{\cal V}}(m_\rho^2) \to \zg \to
\Az \to \tilde M_\rho \to C_{1:3},
(C_{1:3}-C_{2:1}) \to C_{2:1} \to
\Gamma_{{\cal V}\pi\pi}'(0)$ for each sample, with no intermediate
quantity fixed to a central value.  This preserves the correlations
inherited from the common inputs.  Treating $\zg$, $C_{1:3}$ and
$C_{2:1}$ as independent instead would overestimate the uncertainty of
$\Gamma_{{\cal V}\pi\pi}'(0)$ by about a factor of two.

\section{Electromagnetic form factor of the pion\label{sec-ee}}

For the timelike process we define the pion electromagnetic form factor by
\be
\langle \pi^+(q_+)\pi^-(q_-)|j_{\rm em}^\mu(0)|0\rangle
&=&(q_+-q_-)^\mu F_\pi(t),
\qquad
t\equiv(q_++q_-)^2 .
\nonumber\ee
By crossing, the same function describes the usual pion electromagnetic
form factor in the spacelike region.

\begin{figure}
\centering
\begin{subfigure}{0.23\textwidth} \centering
   \begin{tikzpicture}[line width=0.8pt] \begin{feynman}
   \vertex (e); \node [right=1cm of e, blob] (v); 
   \vertex [below right=of v] (a); \vertex [above right=of v] (b);
   \diagram*{
      (e) --[photon] (v); 
      (a) --[scalar] (v) -- [scalar] (b);
   };
   \end{feynman} \end{tikzpicture}
   \caption{} \end{subfigure}
\begin{subfigure}{0.23\textwidth} \centering
   \begin{tikzpicture}[line width=0.8pt] \begin{feynman}
   \vertex (e); \node [right=1cm of e, blob] (v); \node [right=of v, blob] (c);
   \vertex [below right=of c] (a); \vertex [above right=of c] (b);
   \diagram*{
      (e) --[photon] (v) -- [\vmesonfull] (c);
      (a) --[scalar] (c) -- [scalar] (b);
   }; 
   \end{feynman} \end{tikzpicture}
   \caption{} \end{subfigure}
\caption{Diagrams for the electromagnetic form factor of the pion:
(a) the 1PI ${\cal V}\pi\pi$ vertex and (b) $\rho$ exchange.
Wavy, dashed, and double lines denote the external electromagnetic
field, pions, and the $\rho$ meson, respectively; blobs denote
renormalized vertex functions.
\label{Fpi-fig}}
\end{figure}

The pion electromagnetic form factor is the sum of the 1PI
${\cal V}\pi\pi$ vertex and the $\rho$-exchange contribution.  For
on-shell equal-mass pions, $k\cdot(q_a-q_b)=0$, so only the transverse
part of the $\rho$ propagator and of the $\rho{\cal V}$ mixing vertex
contributes.  One then obtains
\be
F_\pi(t) &=&  \Gamma_{{\cal V}\pi\pi}(t) 
\nonumber\\ 
&+&
\frac{\ar}{2}
\frac{\ar \zg^2 f_\pi^2}{\ar \zg^2 f_\pi^2-t A_\rho(t)}
\frac{\Gamma_{\rho\pi\pi}(t)}{\Gamma_{\rho\pi\pi}(0)}
\frac{\Gamma_{\rho{\cal V}}(t)}{\Gamma_{\rho{\cal V}}(0)}.
\label{Fpit}\ee
The factor $\ar/2$ in Eq.~(\ref{Fpit}) is required by charge
normalization: together with
$\Gamma_{{\cal V}\pi\pi}(0)=1-\ar/2$, it gives $F_\pi(0)=1$ for
arbitrary $\ar$.  For $\ar=2$, the small-$t$ expansion
$F_\pi(t)=1+\langle r_\pi^2\rangle t/6+{\cal O}(t^2)$ gives
\be
\frac{1}{6} \langle r_\pi^2\rangle
= 
\Gamma_{{\cal V}\pi\pi}'(0) + \frac{1}{\tilde M_\rho^2}
+ \frac{1}{2 f_\pi^2} \left(2 C_{1:3}- C_{2:1}\right),
\ee
where $\langle r_\pi^2\rangle$ is the square of the charge radius of the
pion and the prime denotes differentiation with respect to $t$.
The experimental value
$\langle r_\pi^2\rangle  = (0.434\pm 0.005)\ \text{fm}^2$~\cite{pdg2026}  
and the values given in Eqs.~(\ref{C13comb}) and (\ref{C21}),
with their correlation retained, lead to
\be
\Gamma_{{\cal V}\pi\pi}(t) &=&
\Gamma_{{\cal V}\pi\pi}'(0)\,t,
\nonumber\\
\Gamma_{{\cal V}\pi\pi}'(0) &=&
\frac{C_{1:2}-C_{2:2}}{2 f_\pi^2}
= (-0.022\pm0.025)\ \text{GeV}^{-2}.
\label{C1222}\ee
Thus, for $\ar=2$, the nonlocal one-loop contribution to the direct
photon--pion vertex, namely the term proportional to $\bar f_3(t)$ in
Eq.~(\ref{Vpipi-finite}), vanishes at NLO.  The accompanying local term proportional to
$\left(1-\ar/2\right)^2$ vanishes at the same time, leaving the finite
local contribution proportional to $C_{1:2}-C_{2:2}$.  Equation~(\ref{C1222})
shows that the pion charge radius constrains this remaining local term
to be consistent with zero within the uncertainties of the inputs.
The choice $\ar=2$ already removes the direct coupling at the origin,
$\Gamma_{{\cal V}\pi\pi}(0)=1-\ar/2$, and this part is an input.  By
contrast, neither the disappearance of the nonlocal loop correction nor
the smallness of the surviving local NLO term is imposed as a
renormalization condition.  Complete VMD is therefore a good numerical
approximation at this order rather than an assumption.

It is worth asking how small this result is compared with its natural
size.  A local term of this kind is expected on dimensional grounds to
be of order $1/\Lambda_\chi^2\simeq1\ \text{GeV}^{-2}$.  The three
contributions to the charge radius are, in these units,
$\langle r_\pi^2\rangle/6=1.859\pm0.023$,
$1/\tilde M_\rho^2=1.700\pm0.001$, and
$(2C_{1:3}-C_{2:1})/(2f_\pi^2)=0.180\pm0.010$.
Their cancellation leaves the value in Eq.~(\ref{C1222}), which is
about forty times smaller than the natural size.  The direct
photon--pion coupling is therefore well below what dimensional
analysis would allow, and the empirical charge radius is saturated by
the $\rho$ to within the present accuracy.  The uncertainty is
dominated by the pion charge radius; the $\rho$-sector input contributes
about half as much.

The value $\ar=2$ is special in this respect.  For general $\ar$ the
direct vertex does not vanish at the origin, $\Gamma_{{\cal V}\pi\pi}(0)
= 1 - \ar/2$, and the form factor can still be normalized and fitted, but
it is then shared between the pole and a direct term of comparable size.
What is found here is that at the KSRF value the direct term is absent at
the origin, that the one-loop correction does not reintroduce a nonlocal
piece, and that the local piece which survives is bounded well below its
natural size.  The separation into direct and $\rho$-pole contributions
is representation dependent; the physical statement concerns the full
form factor $F_\pi(t)$~\cite{Ecker_NPB,Eckerl}.

\begin{figure}
\centering
\begin{subfigure}{.5\textwidth}
  \centering
  \includegraphics[width=.9\linewidth]{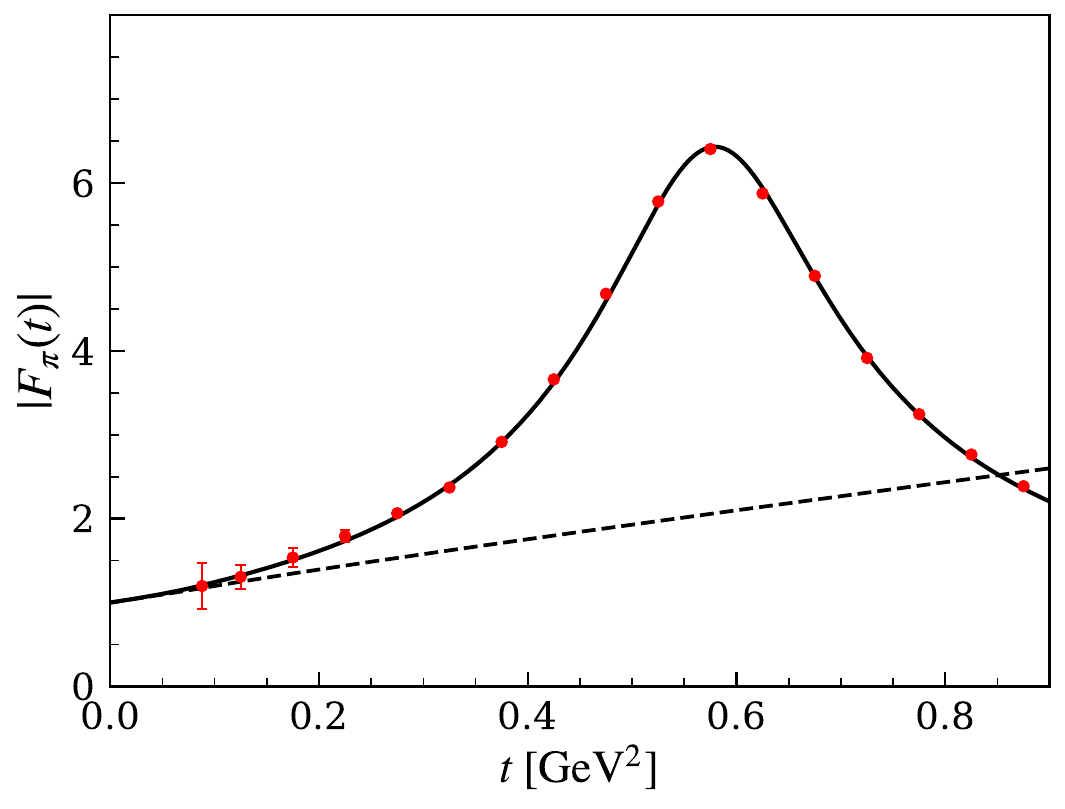}
\end{subfigure}%
\\
\begin{subfigure}{.5\textwidth}
  \centering
  \includegraphics[width=.9\linewidth]{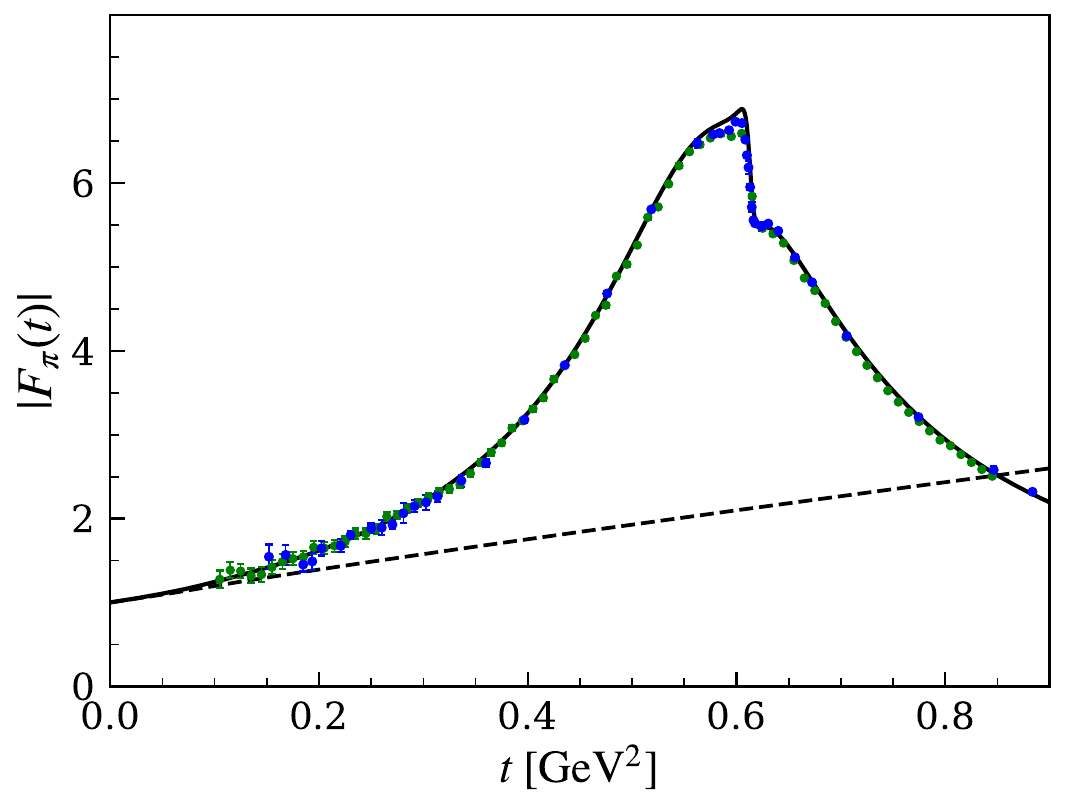}
\end{subfigure}
\caption{
The electromagnetic form factor of the pion, $\left|F_\pi(t)\right|$.
The black solid lines show the present result with $\ar=2$, and the
black dashed lines the vector-meson-less one-loop ChPT result of
Eq.~(\ref{FpiChPT}).  The top panel shows the Belle
$\tau^-\to\pi^-\pi^0\nu_\tau$ data~\cite{Belle} (red), while the
bottom panel shows the KLOE~\cite{KLOE} (green) and SND~\cite{SND}
(blue) $e^+e^-\to\pi^+\pi^-$ data.
}
\label{fig:results}
\end{figure}

The resulting $|F_\pi(t)|$ with $\ar=2$ is compared in
Fig.~\ref{fig:results} with the vector-meson-less one-loop ChPT result
and with the experimental data.  As noted below Eq.~(\ref{Vpipi}), the
$\ar=0$ limit reduces to the standard one-loop ChPT result.  Fixing its
counterterm by the same empirical charge radius gives
\be
F_\pi^{\rm ChPT}(t)
=1+\frac{\langle r_\pi^2\rangle}{6}\,t
+\frac{t}{f_\pi^2}\bar f_3(t).
\label{FpiChPT}\ee
Since $d[t\bar f_3(t)]/dt|_{t=0}=\bar f_3(0)=0$, the two curves have the
same value and slope at the origin, and their difference starts in the
curvature.  No fit to the nonzero-$t$ form-factor data is made.  For
$\ar=2$, the finite LEC combinations are fixed from the pion charge
radius and the $\rho\to\pi\pi$ and $\rho\to e^+e^-$ widths.  The
comparison is made at the same pion-sector chiral order and is intended
to show how the vector-meson-less one-loop ChPT breaks down as $t$
approaches the $\rho$ region.  It should not be interpreted as a proof
of faster order-by-order convergence, because only one nontrivial order
is calculated here and the propagator is partly resummed.

\subsection{Status of the resummation and of the
resonance-region agreement}

The agreement exhibited in Fig.~\ref{fig:results}
extends into the resonance region $t \sim m_\rho^2$,
where the expansion parameter of the present counting
is no longer small,
and it is important to state precisely
what is, and what is not, claimed there.
The dressed propagator of Eq.~(\ref{rho-prop})
re-sums the one-loop self-energy insertions
of the vector meson to all orders.
For $t = {\cal O}(Q^2) \ll \Lambda_\chi^2$
this resummation is harmless:
the difference between the re-summed and the
truncated propagator is of higher chiral order,
and the strict content of the theory
at a given order is unaffected.
At $t \sim m_\rho^2$, in contrast,
the counting itself does not single out
the self-energy chain among the diagrams of the
same nominal order,
and the resummation adopted here is
a choice --- the minimal one that
moves the vector-meson pole off the real axis
and implements the physical decay width
via Eqs.~(\ref{rho-pole},\ref{Arhot}) ---
rather than a consequence of the power counting.
A systematic treatment of the resonance region
requires a dedicated organization of the expansion.  For vector
mesons, the complex-mass scheme provides one such framework
~\cite{cms,gegelia}; analogous resonance countings are used for the
$\Delta(1232)$~\cite{pascalutsa}.  Such an extension is beyond the
scope of this work.
The resonance-region agreement of
Fig.~\ref{fig:results} should therefore be read
as evidence that the $\ar=2$ scheme, with the
self-energy resummation used here, reproduces the
phenomenologically successful VMD form factor ---
which it contains as its
$\Gamma_{{\cal V}\pi\pi}=0$ limit ---
and not as a controlled prediction of the
expansion in that region.

An important point is that this construction is not simply a model
obtained by adding an explicit $\rho$ pole to ChPT.  Because the vector
mesons are introduced within a chiral Lagrangian with a definite power
counting, the low-energy expansion is constrained from the outset.  At
any fixed order, only a finite set of additional LEC combinations can
enter, and the low-momentum expansion of the result agrees with the
renormalized EFT through that order.  The $\rho$-pole observables used
to determine some of these LECs therefore do not introduce arbitrary
momentum dependence into the low-energy form factor.

The continuation into the resonance region is constrained by the same
low-energy theory.  When expanded for $t\ll m_\rho^2$, the resummed
expression reproduces the NLO result derived above, with differences
only at higher chiral order.  It may therefore be viewed as a
phenomenological continuation of the low-energy EFT into the resonance
region, rather than as an independent model.

Recent $e^+e^-\to\pi^+\pi^-$ measurements by BABAR, BESIII, and
CMD-3 provide additional high-precision information in this region
~\cite{BaBar2012,BESIII2016,CMD32024}.  These data sets are not fully
consistent with one another in the $\rho$ region
~\cite{Davier2024}.  Since the present work does not fit the
form-factor data, Fig.~\ref{fig:results} keeps KLOE and SND as
representative data sets.  A precision analysis of the resonance
region would have to address these data tensions together with a more
systematic treatment of isospin breaking and radiative corrections.

\subsection{$\rho-\omega$ mixing}
The form factor extracted from $e^+e^-\to\pi^+\pi^-$, unlike the
isovector form factor from $\tau^-\to\pi^-\pi^0\nu_\tau$, receives a
visible $\rho$--$\omega$ mixing contribution~\cite{thomas,urec,gegelia}.
Following Refs.~\cite{qmass,urec}, we multiply the $\rho$-exchange term
in Eq.~(\ref{Fpit}) by
\be
1 + \frac{F_\omega}{F_\rho}
\frac{\Theta_{\rho\omega}}{t - m_\omega^2 +
i m_\omega\Gamma_\omega},
\ee
Near the $\rho$--$\omega$ region,
$\Theta_{\rho\omega}=2m_\rho(m_u-m_d)+e^2F_\rho F_\omega$.
The vector-meson couplings are defined by
$\langle0|j_\mu^{\text{em}}|\rho^0\rangle=\varepsilon_\mu F_\rho m_\rho$
and
$\langle0|j_\mu^{\text{em}}|\omega\rangle=\varepsilon_\mu F_\omega m_\omega$,
where $j_\mu^{\text{em}}$ is the electromagnetic current,
$\varepsilon_\mu$ is the polarization vector, and $e^2=4\pi\alpha$.
With the values estimated in \cite{urec}, ${F_\omega}/{F_\rho}=\frac13$,
$\Theta_{\rho\omega}= (-3.91\pm0.30)\times 10^{-3}\ \text{GeV}^2$,
and with the experimental value~\cite{pdg2026}
$m_\omega=782.66\ \text{MeV}$ and $\Gamma_\omega=8.68\ \text{MeV}$,
the resulting pion form factor is drawn in
the bottom panel of Fig.~\ref{fig:results}.  Other isospin-breaking
and electromagnetic corrections that are needed for a precision
comparison of $\tau$ and $e^+e^-$ data are not included here; the
bottom panel is intended only as a phenomenological illustration.

\section{Summary\label{sec-summary}}
We have developed a version of ChPT in which vector mesons are kept as
explicit intermediate degrees of freedom.  Their masses are counted as
${\cal O}(\Lambda_\chi)$ and the momenta they carry as ${\cal O}(Q)$.
The vector-meson kinetic term therefore starts at NLO, and the LO
propagator is momentum independent.

At one loop, the divergences generated by the LO Lagrangian are
absorbed by the NLO counterterms, giving a consistent renormalized
expansion at this order.  We applied the formalism to the pion
electromagnetic form factor.  For the KSRF value $\ar=2$, the nonlocal
one-loop correction to the direct photon--pion vertex vanishes at NLO.
A finite local term remains, but the empirical pion charge radius gives
a value consistent with zero and well below its natural size.  For
$\ar=2$, the direct photon--pion contribution is therefore very small,
and the form factor is largely saturated by $\rho$ exchange, which
corresponds to VMD.

At low momentum transfer the NLO-dressed $\rho$ propagator, kept in its
resummed form, reproduces the observed curvature more efficiently than
the vector-meson-less one-loop result.  This observation is
encouraging, but with only one nontrivial order it should not be read as
a quantitative demonstration of faster EFT convergence.

We have also continued the result into the $\rho$ region using a
resummation of the one-loop $\rho$ self-energy.  The resonance
region itself lies outside the strict counting of this work, while the
continuation reproduces the NLO low-energy expansion by construction.

The counting used here is designed for low-energy processes in which
vector mesons appear as intermediate states with momenta small compared
with $m_V$.  It keeps the low-energy expansion controlled while
retaining the vector mesons explicitly, and is particularly suited to
applications such as nuclear forces and nuclear electroweak currents.
The present one-loop mesonic calculation provides the renormalized
building blocks for extending this framework to the nucleon sector.
That extension is carried out in a companion paper~\cite{ParkNN2026}.

\section*{Acknowledgments}
The author 
thanks Prof. S.-W. Hong and Prof. A.W. Thomas
for useful discussions and their support.
He is also grateful to Dr. Joochun (Jason) Park 
for his careful reading of the manuscript and helpful comments.
This work was supported by the IBS grant funded by the Korea government (No. IBS-R031-D1).

\appendix

\section{Loop function\label{loop-ftn}}

The loop function $f_3(k^2)$ is defined as~\cite{tsp-report}, with
$\varepsilon=(4-d)/2$,
\be
f_3(k^2) = \frac{\Gamma(\varepsilon)}{16\pi^2}
\int_0^1\!\!dz\, z (2z-1) \left(
\frac{\zm^2 - z(1-z) k^2}{4\pi\mu^2} \right)^{-\varepsilon}.
\label{f3-def}\ee
At $k^2=0$, it is
\be
f_3(0) = -\frac16 \left[
   2 \lambda + \frac{1}{16\pi^2} \left(\ln\frac{m_\pi^2}{\mu^2}-1\right)\right].
\ee
After subtraction the loop function is finite.  At NLO we may then
replace $\zm$ by the physical $m_\pi$; the difference is of higher
order.  With this replacement,
$\bar f_3(k^2)\equiv f_3(k^2)-f_3(0)$ reads
\be
\bar f_3(k^2) &=& 
\frac{1}{16\pi^2} \int_{2 m_\pi}^\infty\!\!dM\, \frac{1}{3M}
\left(1-\frac{4m_\pi^2}{M^2} \right)^{3/2}
\nonumber\\
&&\times\
\frac{k^2}{M^2-k^2-i0}
\nonumber \\
&=& \frac{1}{16\pi^2} \left[
\frac{1}{9} + \frac{1}{3 \sigma^3}
\left(\arctan{\sigma} -\sigma\right)
\right].
\label{f3bar-def}\ee
where
$\sigma \equiv {k}/{\sqrt{4 m_\pi^2-k^2-i0}}$.
Its imaginary part reads
\be
\Im f_3(k^2) = 
\frac{1}{96\pi} 
\left(\frac{k^2-4m_\pi^2}{k^2}\right)^{3/2}
\theta(k^2-4m_\pi^2),
\ee
where $\theta$ is the Heaviside step function, and
$\bar f_3(m_\rho^2)  
= (-0.340 + 2.692i)\times 10^{-3}$
and $m_\rho^2 f_3'(m_\rho^2)=  (-1.289 + 0.602i)\times 10^{-3}$.


\begin{thebibliography}{51}%
\makeatletter
\providecommand \@ifxundefined [1]{%
 \@ifx{#1\undefined}
}%
\providecommand \@ifnum [1]{%
 \ifnum #1\expandafter \@firstoftwo
 \else \expandafter \@secondoftwo
 \fi
}%
\providecommand \@ifx [1]{%
 \ifx #1\expandafter \@firstoftwo
 \else \expandafter \@secondoftwo
 \fi
}%
\providecommand \natexlab [1]{#1}%
\providecommand \enquote  [1]{``#1''}%
\providecommand \bibnamefont  [1]{#1}%
\providecommand \bibfnamefont [1]{#1}%
\providecommand \citenamefont [1]{#1}%
\providecommand \href@noop [0]{\@secondoftwo}%
\providecommand \href [0]{\begingroup \@sanitize@url \@href}%
\providecommand \@href[1]{\@@startlink{#1}\@@href}%
\providecommand \@@href[1]{\endgroup#1\@@endlink}%
\providecommand \@sanitize@url [0]{\catcode `\\12\catcode `\$12\catcode
  `\&12\catcode `\#12\catcode `\^12\catcode `\_12\catcode `\%12\relax}%
\providecommand \@@startlink[1]{}%
\providecommand \@@endlink[0]{}%
\providecommand \url  [0]{\begingroup\@sanitize@url \@url }%
\providecommand \@url [1]{\endgroup\@href {#1}{\urlprefix }}%
\providecommand \urlprefix  [0]{URL }%
\providecommand \Eprint [0]{\href }%
\providecommand \doibase [0]{https://doi.org/}%
\providecommand \selectlanguage [0]{\@gobble}%
\providecommand \bibinfo  [0]{\@secondoftwo}%
\providecommand \bibfield  [0]{\@secondoftwo}%
\providecommand \translation [1]{[#1]}%
\providecommand \BibitemOpen [0]{}%
\providecommand \bibitemStop [0]{}%
\providecommand \bibitemNoStop [0]{.\EOS\space}%
\providecommand \EOS [0]{\spacefactor3000\relax}%
\providecommand \BibitemShut  [1]{\csname bibitem#1\endcsname}%
\let\auto@bib@innerbib\@empty
\bibitem [{\citenamefont {Weinberg}(1968)}]{wein68}%
  \BibitemOpen
  \bibfield  {author} {\bibinfo {author} {\bibfnamefont {S.}~\bibnamefont
  {Weinberg}},\ }\href@noop {} {\bibfield  {journal} {\bibinfo  {journal}
  {Physical Review}\ }\textbf {\bibinfo {volume} {166}},\ \bibinfo {pages}
  {1568} (\bibinfo {year} {1968})}\BibitemShut {NoStop}%
\bibitem [{\citenamefont {Coleman}\ \emph {et~al.}(1969)\citenamefont
  {Coleman}, \citenamefont {Wess},\ and\ \citenamefont {Zumino}}]{ccwz1}%
  \BibitemOpen
  \bibfield  {author} {\bibinfo {author} {\bibfnamefont {S.}~\bibnamefont
  {Coleman}}, \bibinfo {author} {\bibfnamefont {J.}~\bibnamefont {Wess}},\ and\
  \bibinfo {author} {\bibfnamefont {B.}~\bibnamefont {Zumino}},\ }\href@noop {}
  {\bibfield  {journal} {\bibinfo  {journal} {Physical Review}\ }\textbf
  {\bibinfo {volume} {177}},\ \bibinfo {pages} {2239} (\bibinfo {year}
  {1969})}\BibitemShut {NoStop}%
\bibitem [{\citenamefont {Callan~Jr}\ \emph {et~al.}(1969)\citenamefont
  {Callan~Jr}, \citenamefont {Coleman}, \citenamefont {Wess},\ and\
  \citenamefont {Zumino}}]{ccwz}%
  \BibitemOpen
  \bibfield  {author} {\bibinfo {author} {\bibfnamefont {C.~G.}\ \bibnamefont
  {Callan~Jr}}, \bibinfo {author} {\bibfnamefont {S.}~\bibnamefont {Coleman}},
  \bibinfo {author} {\bibfnamefont {J.}~\bibnamefont {Wess}},\ and\ \bibinfo
  {author} {\bibfnamefont {B.}~\bibnamefont {Zumino}},\ }\href@noop {}
  {\bibfield  {journal} {\bibinfo  {journal} {Physical Review}\ }\textbf
  {\bibinfo {volume} {177}},\ \bibinfo {pages} {2247} (\bibinfo {year}
  {1969})}\BibitemShut {NoStop}%
\bibitem [{\citenamefont {Gasser}\ and\ \citenamefont
  {Leutwyler}(1984)}]{GL84}%
  \BibitemOpen
  \bibfield  {author} {\bibinfo {author} {\bibfnamefont {J.}~\bibnamefont
  {Gasser}}\ and\ \bibinfo {author} {\bibfnamefont {H.}~\bibnamefont
  {Leutwyler}},\ }\href@noop {} {\bibfield  {journal} {\bibinfo  {journal}
  {Annals of Physics}\ }\textbf {\bibinfo {volume} {158}},\ \bibinfo {pages}
  {142} (\bibinfo {year} {1984})}\BibitemShut {NoStop}%
\bibitem [{\citenamefont {Gasser}\ and\ \citenamefont
  {Leutwyler}(1985)}]{GL85}%
  \BibitemOpen
  \bibfield  {author} {\bibinfo {author} {\bibfnamefont {J.}~\bibnamefont
  {Gasser}}\ and\ \bibinfo {author} {\bibfnamefont {H.}~\bibnamefont
  {Leutwyler}},\ }\href@noop {} {\bibfield  {journal} {\bibinfo  {journal}
  {Nuclear Physics B}\ }\textbf {\bibinfo {volume} {250}},\ \bibinfo {pages}
  {465} (\bibinfo {year} {1985})}\BibitemShut {NoStop}%
\bibitem [{\citenamefont {Weinberg}(1990)}]{weinberg-counting}%
  \BibitemOpen
  \bibfield  {author} {\bibinfo {author} {\bibfnamefont {S.}~\bibnamefont
  {Weinberg}},\ }\href@noop {} {\bibfield  {journal} {\bibinfo  {journal}
  {Physics Letters,(Section) B;(Netherlands)}\ }\textbf {\bibinfo {volume}
  {251}} (\bibinfo {year} {1990})}\BibitemShut {NoStop}%
\bibitem [{\citenamefont {Weinberg}(1991)}]{weinberg-counting2}%
  \BibitemOpen
  \bibfield  {author} {\bibinfo {author} {\bibfnamefont {S.}~\bibnamefont
  {Weinberg}},\ }\href@noop {} {\bibfield  {journal} {\bibinfo  {journal}
  {Nuclear Physics B}\ }\textbf {\bibinfo {volume} {363}},\ \bibinfo {pages}
  {3} (\bibinfo {year} {1991})}\BibitemShut {NoStop}%
\bibitem [{\citenamefont {Krebs}\ \emph {et~al.}(2007)\citenamefont {Krebs},
  \citenamefont {Epelbaum},\ and\ \citenamefont {Mei{\ss}ner}}]{epj32}%
  \BibitemOpen
  \bibfield  {author} {\bibinfo {author} {\bibfnamefont {H.}~\bibnamefont
  {Krebs}}, \bibinfo {author} {\bibfnamefont {E.}~\bibnamefont {Epelbaum}},\
  and\ \bibinfo {author} {\bibfnamefont {U.~G.}\ \bibnamefont {Mei{\ss}ner}},\
  }\href@noop {} {\bibfield  {journal} {\bibinfo  {journal} {The European
  Physical Journal A}\ }\textbf {\bibinfo {volume} {32}},\ \bibinfo {pages}
  {127} (\bibinfo {year} {2007})}\BibitemShut {NoStop}%
\bibitem [{\citenamefont {Jiang}\ \emph {et~al.}(2020)\citenamefont {Jiang},
  \citenamefont {Ekstr{\"o}m}, \citenamefont {Forss{\'e}n}, \citenamefont
  {Hagen}, \citenamefont {Jansen},\ and\ \citenamefont {Papenbrock}}]{prc102}%
  \BibitemOpen
  \bibfield  {author} {\bibinfo {author} {\bibfnamefont {W.}~\bibnamefont
  {Jiang}}, \bibinfo {author} {\bibfnamefont {A.}~\bibnamefont {Ekstr{\"o}m}},
  \bibinfo {author} {\bibfnamefont {C.}~\bibnamefont {Forss{\'e}n}}, \bibinfo
  {author} {\bibfnamefont {G.}~\bibnamefont {Hagen}}, \bibinfo {author}
  {\bibfnamefont {G.}~\bibnamefont {Jansen}},\ and\ \bibinfo {author}
  {\bibfnamefont {T.}~\bibnamefont {Papenbrock}},\ }\href@noop {} {\bibfield
  {journal} {\bibinfo  {journal} {Physical Review C}\ }\textbf {\bibinfo
  {volume} {102}},\ \bibinfo {pages} {054301} (\bibinfo {year}
  {2020})}\BibitemShut {NoStop}%
\bibitem [{\citenamefont {Nosyk}\ \emph {et~al.}(2021)\citenamefont {Nosyk},
  \citenamefont {Entem},\ and\ \citenamefont {Machleidt}}]{prc104}%
  \BibitemOpen
  \bibfield  {author} {\bibinfo {author} {\bibfnamefont {Y.}~\bibnamefont
  {Nosyk}}, \bibinfo {author} {\bibfnamefont {D.}~\bibnamefont {Entem}},\ and\
  \bibinfo {author} {\bibfnamefont {R.}~\bibnamefont {Machleidt}},\ }\href@noop
  {} {\bibfield  {journal} {\bibinfo  {journal} {Physical Review C}\ }\textbf
  {\bibinfo {volume} {104}},\ \bibinfo {pages} {054001} (\bibinfo {year}
  {2021})}\BibitemShut {NoStop}%
\bibitem [{\citenamefont {Ecker}\ \emph
  {et~al.}(1989{\natexlab{a}})\citenamefont {Ecker}, \citenamefont {Gasser},
  \citenamefont {Pich},\ and\ \citenamefont {De~Rafael}}]{Ecker_NPB}%
  \BibitemOpen
  \bibfield  {author} {\bibinfo {author} {\bibfnamefont {G.}~\bibnamefont
  {Ecker}}, \bibinfo {author} {\bibfnamefont {J.}~\bibnamefont {Gasser}},
  \bibinfo {author} {\bibfnamefont {A.}~\bibnamefont {Pich}},\ and\ \bibinfo
  {author} {\bibfnamefont {E.}~\bibnamefont {De~Rafael}},\ }\href@noop {}
  {\bibfield  {journal} {\bibinfo  {journal} {Nuclear Physics B}\ }\textbf
  {\bibinfo {volume} {321}},\ \bibinfo {pages} {311} (\bibinfo {year}
  {1989}{\natexlab{a}})}\BibitemShut {NoStop}%
\bibitem [{\citenamefont {Ecker}\ \emph
  {et~al.}(1989{\natexlab{b}})\citenamefont {Ecker}, \citenamefont {Gasser},
  \citenamefont {Leutwyler}, \citenamefont {Pich},\ and\ \citenamefont
  {De~Rafael}}]{Eckerl}%
  \BibitemOpen
  \bibfield  {author} {\bibinfo {author} {\bibfnamefont {G.}~\bibnamefont
  {Ecker}}, \bibinfo {author} {\bibfnamefont {J.}~\bibnamefont {Gasser}},
  \bibinfo {author} {\bibfnamefont {H.}~\bibnamefont {Leutwyler}}, \bibinfo
  {author} {\bibfnamefont {A.}~\bibnamefont {Pich}},\ and\ \bibinfo {author}
  {\bibfnamefont {E.}~\bibnamefont {De~Rafael}},\ }\href@noop {} {\bibfield
  {journal} {\bibinfo  {journal} {Physics Letters B}\ }\textbf {\bibinfo
  {volume} {223}},\ \bibinfo {pages} {425} (\bibinfo {year}
  {1989}{\natexlab{b}})}\BibitemShut {NoStop}%
\bibitem [{\citenamefont {Georgi}(1990)}]{georgi1990}%
  \BibitemOpen
  \bibfield  {author} {\bibinfo {author} {\bibfnamefont {H.}~\bibnamefont
  {Georgi}},\ }\href@noop {} {\bibfield  {journal} {\bibinfo  {journal}
  {Nuclear Physics B}\ }\textbf {\bibinfo {volume} {331}},\ \bibinfo {pages}
  {311} (\bibinfo {year} {1990})}\BibitemShut {NoStop}%
\bibitem [{\citenamefont {Bando}\ \emph {et~al.}(1985)\citenamefont {Bando},
  \citenamefont {Kugo}, \citenamefont {Uehara}, \citenamefont {Yamawaki},\ and\
  \citenamefont {Yanagida}}]{bando}%
  \BibitemOpen
  \bibfield  {author} {\bibinfo {author} {\bibfnamefont {M.}~\bibnamefont
  {Bando}}, \bibinfo {author} {\bibfnamefont {T.}~\bibnamefont {Kugo}},
  \bibinfo {author} {\bibfnamefont {S.}~\bibnamefont {Uehara}}, \bibinfo
  {author} {\bibfnamefont {K.}~\bibnamefont {Yamawaki}},\ and\ \bibinfo
  {author} {\bibfnamefont {T.}~\bibnamefont {Yanagida}},\ }\href@noop {}
  {\bibfield  {journal} {\bibinfo  {journal} {Physical Review Letters}\
  }\textbf {\bibinfo {volume} {54}},\ \bibinfo {pages} {1215} (\bibinfo {year}
  {1985})}\BibitemShut {NoStop}%
\bibitem [{\citenamefont {Bando}\ \emph {et~al.}(1988)\citenamefont {Bando},
  \citenamefont {Kugo},\ and\ \citenamefont {Yamawaki}}]{bando-review}%
  \BibitemOpen
  \bibfield  {author} {\bibinfo {author} {\bibfnamefont {M.}~\bibnamefont
  {Bando}}, \bibinfo {author} {\bibfnamefont {T.}~\bibnamefont {Kugo}},\ and\
  \bibinfo {author} {\bibfnamefont {K.}~\bibnamefont {Yamawaki}},\ }\href
  {https://doi.org/10.1016/0370-1573(88)90019-1} {\bibfield  {journal}
  {\bibinfo  {journal} {Physics Reports}\ }\textbf {\bibinfo {volume} {164}},\
  \bibinfo {pages} {217} (\bibinfo {year} {1988})}\BibitemShut {NoStop}%
\bibitem [{\citenamefont {Harada}\ and\ \citenamefont {Yamawaki}(2003)}]{hls}%
  \BibitemOpen
  \bibfield  {author} {\bibinfo {author} {\bibfnamefont {M.}~\bibnamefont
  {Harada}}\ and\ \bibinfo {author} {\bibfnamefont {K.}~\bibnamefont
  {Yamawaki}},\ }\href@noop {} {\bibfield  {journal} {\bibinfo  {journal}
  {Physics Reports}\ }\textbf {\bibinfo {volume} {381}},\ \bibinfo {pages} {1}
  (\bibinfo {year} {2003})}\BibitemShut {NoStop}%
\bibitem [{\citenamefont {Harada}\ and\ \citenamefont
  {Yamawaki}(1992)}]{HaradaYamawaki1992}%
  \BibitemOpen
  \bibfield  {author} {\bibinfo {author} {\bibfnamefont {M.}~\bibnamefont
  {Harada}}\ and\ \bibinfo {author} {\bibfnamefont {K.}~\bibnamefont
  {Yamawaki}},\ }\href {https://doi.org/10.1016/0370-2693(92)91084-M}
  {\bibfield  {journal} {\bibinfo  {journal} {Phys. Lett. B}\ }\textbf
  {\bibinfo {volume} {297}},\ \bibinfo {pages} {151} (\bibinfo {year}
  {1992})},\ \Eprint {https://arxiv.org/abs/hep-ph/9210208}
  {arXiv:hep-ph/9210208 [hep-ph]} \BibitemShut {NoStop}%
\bibitem [{\citenamefont {Jenkins}\ \emph {et~al.}(1995)\citenamefont
  {Jenkins}, \citenamefont {Manohar},\ and\ \citenamefont {Wise}}]{jmw}%
  \BibitemOpen
  \bibfield  {author} {\bibinfo {author} {\bibfnamefont {E.}~\bibnamefont
  {Jenkins}}, \bibinfo {author} {\bibfnamefont {A.~V.}\ \bibnamefont
  {Manohar}},\ and\ \bibinfo {author} {\bibfnamefont {M.~B.}\ \bibnamefont
  {Wise}},\ }\href@noop {} {\bibfield  {journal} {\bibinfo  {journal} {Physical
  Review Letters}\ }\textbf {\bibinfo {volume} {75}},\ \bibinfo {pages} {2272}
  (\bibinfo {year} {1995})}\BibitemShut {NoStop}%
\bibitem [{\citenamefont {Bijnens}\ \emph {et~al.}(1997)\citenamefont
  {Bijnens}, \citenamefont {Gosdzinsky},\ and\ \citenamefont {Talavera}}]{bgt}%
  \BibitemOpen
  \bibfield  {author} {\bibinfo {author} {\bibfnamefont {J.}~\bibnamefont
  {Bijnens}}, \bibinfo {author} {\bibfnamefont {P.}~\bibnamefont
  {Gosdzinsky}},\ and\ \bibinfo {author} {\bibfnamefont {P.}~\bibnamefont
  {Talavera}},\ }\href@noop {} {\bibfield  {journal} {\bibinfo  {journal}
  {Nuclear Physics B}\ }\textbf {\bibinfo {volume} {501}},\ \bibinfo {pages}
  {495} (\bibinfo {year} {1997})}\BibitemShut {NoStop}%
\bibitem [{\citenamefont {Rosell}\ \emph {et~al.}(2004)\citenamefont {Rosell},
  \citenamefont {Sanz-Cillero},\ and\ \citenamefont {Pich}}]{rxt-loop}%
  \BibitemOpen
  \bibfield  {author} {\bibinfo {author} {\bibfnamefont {I.}~\bibnamefont
  {Rosell}}, \bibinfo {author} {\bibfnamefont {J.~J.}\ \bibnamefont
  {Sanz-Cillero}},\ and\ \bibinfo {author} {\bibfnamefont {A.}~\bibnamefont
  {Pich}},\ }\href@noop {} {\bibfield  {journal} {\bibinfo  {journal} {Journal
  of High Energy Physics}\ }\textbf {\bibinfo {volume} {2004}},\ \bibinfo
  {pages} {042} (\bibinfo {year} {2004})}\BibitemShut {NoStop}%
\bibitem [{\citenamefont {Djukanovic}\ \emph {et~al.}(2009)\citenamefont
  {Djukanovic}, \citenamefont {Gegelia}, \citenamefont {Keller},\ and\
  \citenamefont {Scherer}}]{cms}%
  \BibitemOpen
  \bibfield  {author} {\bibinfo {author} {\bibfnamefont {D.}~\bibnamefont
  {Djukanovic}}, \bibinfo {author} {\bibfnamefont {J.}~\bibnamefont {Gegelia}},
  \bibinfo {author} {\bibfnamefont {A.}~\bibnamefont {Keller}},\ and\ \bibinfo
  {author} {\bibfnamefont {S.}~\bibnamefont {Scherer}},\ }\href@noop {}
  {\bibfield  {journal} {\bibinfo  {journal} {Physics Letters B}\ }\textbf
  {\bibinfo {volume} {680}},\ \bibinfo {pages} {235} (\bibinfo {year}
  {2009})}\BibitemShut {NoStop}%
\bibitem [{\citenamefont {Djukanovic}\ \emph {et~al.}(2015)\citenamefont
  {Djukanovic}, \citenamefont {Gegelia}, \citenamefont {Keller}, \citenamefont
  {Scherer},\ and\ \citenamefont {Tiator}}]{gegelia}%
  \BibitemOpen
  \bibfield  {author} {\bibinfo {author} {\bibfnamefont {D.}~\bibnamefont
  {Djukanovic}}, \bibinfo {author} {\bibfnamefont {J.}~\bibnamefont {Gegelia}},
  \bibinfo {author} {\bibfnamefont {A.}~\bibnamefont {Keller}}, \bibinfo
  {author} {\bibfnamefont {S.}~\bibnamefont {Scherer}},\ and\ \bibinfo {author}
  {\bibfnamefont {L.}~\bibnamefont {Tiator}},\ }\href@noop {} {\bibfield
  {journal} {\bibinfo  {journal} {Physics Letters B}\ }\textbf {\bibinfo
  {volume} {742}},\ \bibinfo {pages} {55} (\bibinfo {year} {2015})}\BibitemShut
  {NoStop}%
\bibitem [{\citenamefont {Lutz}\ and\ \citenamefont
  {Leupold}(2008)}]{lutz-leupold}%
  \BibitemOpen
  \bibfield  {author} {\bibinfo {author} {\bibfnamefont {M.~F.}\ \bibnamefont
  {Lutz}}\ and\ \bibinfo {author} {\bibfnamefont {S.}~\bibnamefont {Leupold}},\
  }\href@noop {} {\bibfield  {journal} {\bibinfo  {journal} {Nuclear Physics
  A}\ }\textbf {\bibinfo {volume} {813}},\ \bibinfo {pages} {96} (\bibinfo
  {year} {2008})}\BibitemShut {NoStop}%
\bibitem [{\citenamefont {Terschl{\"u}sen}\ and\ \citenamefont
  {Leupold}(2016)}]{terschluesen}%
  \BibitemOpen
  \bibfield  {author} {\bibinfo {author} {\bibfnamefont {C.}~\bibnamefont
  {Terschl{\"u}sen}}\ and\ \bibinfo {author} {\bibfnamefont {S.}~\bibnamefont
  {Leupold}},\ }\href@noop {} {\bibfield  {journal} {\bibinfo  {journal}
  {Physical Review D}\ }\textbf {\bibinfo {volume} {94}},\ \bibinfo {pages}
  {014021} (\bibinfo {year} {2016})}\BibitemShut {NoStop}%
\bibitem [{\citenamefont {Park}\ \emph {et~al.}(1993)\citenamefont {Park},
  \citenamefont {Min},\ and\ \citenamefont {Rho}}]{tsp-report}%
  \BibitemOpen
  \bibfield  {author} {\bibinfo {author} {\bibfnamefont {T.-S.}\ \bibnamefont
  {Park}}, \bibinfo {author} {\bibfnamefont {D.-P.}\ \bibnamefont {Min}},\ and\
  \bibinfo {author} {\bibfnamefont {M.}~\bibnamefont {Rho}},\ }\href@noop {}
  {\bibfield  {journal} {\bibinfo  {journal} {Physics Reports}\ }\textbf
  {\bibinfo {volume} {233}},\ \bibinfo {pages} {341} (\bibinfo {year}
  {1993})}\BibitemShut {NoStop}%
\bibitem [{\citenamefont {Rho}(1991)}]{mrho-counting}%
  \BibitemOpen
  \bibfield  {author} {\bibinfo {author} {\bibfnamefont {M.}~\bibnamefont
  {Rho}},\ }\href@noop {} {\bibfield  {journal} {\bibinfo  {journal} {Physical
  review letters}\ }\textbf {\bibinfo {volume} {66}},\ \bibinfo {pages} {1275}
  (\bibinfo {year} {1991})}\BibitemShut {NoStop}%
\bibitem [{\citenamefont {Bijnens}\ \emph {et~al.}(1998)\citenamefont
  {Bijnens}, \citenamefont {Colangelo},\ and\ \citenamefont
  {Talavera}}]{BCT1998}%
  \BibitemOpen
  \bibfield  {author} {\bibinfo {author} {\bibfnamefont {J.}~\bibnamefont
  {Bijnens}}, \bibinfo {author} {\bibfnamefont {G.}~\bibnamefont {Colangelo}},\
  and\ \bibinfo {author} {\bibfnamefont {P.}~\bibnamefont {Talavera}},\ }\href
  {https://doi.org/10.1088/1126-6708/1998/05/014} {\bibfield  {journal}
  {\bibinfo  {journal} {Journal of High Energy Physics}\ }\textbf {\bibinfo
  {volume} {1998}},\ \bibinfo {pages} {014} (\bibinfo {year} {1998})},\ \Eprint
  {https://arxiv.org/abs/hep-ph/9805389} {arXiv:hep-ph/9805389} \BibitemShut
  {NoStop}%
\bibitem [{\citenamefont {Bijnens}\ and\ \citenamefont
  {Talavera}(2002)}]{BT2002}%
  \BibitemOpen
  \bibfield  {author} {\bibinfo {author} {\bibfnamefont {J.}~\bibnamefont
  {Bijnens}}\ and\ \bibinfo {author} {\bibfnamefont {P.}~\bibnamefont
  {Talavera}},\ }\href {https://doi.org/10.1088/1126-6708/2002/03/046}
  {\bibfield  {journal} {\bibinfo  {journal} {Journal of High Energy Physics}\
  }\textbf {\bibinfo {volume} {2002}},\ \bibinfo {pages} {046} (\bibinfo {year}
  {2002})},\ \Eprint {https://arxiv.org/abs/hep-ph/0203049}
  {arXiv:hep-ph/0203049} \BibitemShut {NoStop}%
\bibitem [{\citenamefont {Gell-Mann}\ and\ \citenamefont
  {Zachariasen}(1961)}]{gell-mann}%
  \BibitemOpen
  \bibfield  {author} {\bibinfo {author} {\bibfnamefont {M.}~\bibnamefont
  {Gell-Mann}}\ and\ \bibinfo {author} {\bibfnamefont {F.}~\bibnamefont
  {Zachariasen}},\ }\href@noop {} {\bibfield  {journal} {\bibinfo  {journal}
  {Physical Review}\ }\textbf {\bibinfo {volume} {124}},\ \bibinfo {pages}
  {953} (\bibinfo {year} {1961})}\BibitemShut {NoStop}%
\bibitem [{\citenamefont {Sakurai}(1969)}]{sakurai}%
  \BibitemOpen
  \bibfield  {author} {\bibinfo {author} {\bibfnamefont {J.~J.}\ \bibnamefont
  {Sakurai}},\ }\href@noop {} {\emph {\bibinfo {title} {Currents and mesons}}}\
  (\bibinfo  {publisher} {University of Chicago press},\ \bibinfo {year}
  {1969})\BibitemShut {NoStop}%
\bibitem [{\citenamefont {Witten}(1979)}]{witten1979baryons}%
  \BibitemOpen
  \bibfield  {author} {\bibinfo {author} {\bibfnamefont {E.}~\bibnamefont
  {Witten}},\ }\href@noop {} {\bibfield  {journal} {\bibinfo  {journal}
  {Nuclear Physics B}\ }\textbf {\bibinfo {volume} {160}},\ \bibinfo {pages}
  {57} (\bibinfo {year} {1979})}\BibitemShut {NoStop}%
\bibitem [{\citenamefont {Zweig}(1964)}]{zweig}%
  \BibitemOpen
  \bibfield  {author} {\bibinfo {author} {\bibfnamefont {G.}~\bibnamefont
  {Zweig}},\ }\href {https://doi.org/10.17181/CERN-TH-412} {\emph {\bibinfo
  {title} {An {SU(3)} Model for Strong Interaction Symmetry and Its
  Breaking}}},\ \bibinfo {type} {Tech. Rep.}\ (\bibinfo  {institution} {CERN},\
  \bibinfo {year} {1964})\BibitemShut {NoStop}%
\bibitem [{\citenamefont {Becher}\ and\ \citenamefont
  {Leutwyler}(1999)}]{infra}%
  \BibitemOpen
  \bibfield  {author} {\bibinfo {author} {\bibfnamefont {T.}~\bibnamefont
  {Becher}}\ and\ \bibinfo {author} {\bibfnamefont {H.}~\bibnamefont
  {Leutwyler}},\ }\href@noop {} {\bibfield  {journal} {\bibinfo  {journal} {The
  European Physical Journal C-Particles and Fields}\ }\textbf {\bibinfo
  {volume} {9}},\ \bibinfo {pages} {643} (\bibinfo {year} {1999})}\BibitemShut
  {NoStop}%
\bibitem [{\citenamefont {Bruns}\ and\ \citenamefont
  {Mei{\ss}ner}(2005)}]{bruns2005}%
  \BibitemOpen
  \bibfield  {author} {\bibinfo {author} {\bibfnamefont {P.~C.}\ \bibnamefont
  {Bruns}}\ and\ \bibinfo {author} {\bibfnamefont {U.-G.}\ \bibnamefont
  {Mei{\ss}ner}},\ }\href {https://doi.org/10.1140/epjc/s2005-02118-0}
  {\bibfield  {journal} {\bibinfo  {journal} {European Physical Journal C}\
  }\textbf {\bibinfo {volume} {40}},\ \bibinfo {pages} {97} (\bibinfo {year}
  {2005})},\ \Eprint {https://arxiv.org/abs/hep-ph/0411223}
  {arXiv:hep-ph/0411223} \BibitemShut {NoStop}%
\bibitem [{\citenamefont {Fuchs}\ \emph {et~al.}(2003)\citenamefont {Fuchs},
  \citenamefont {Gegelia}, \citenamefont {Japaridze},\ and\ \citenamefont
  {Scherer}}]{extended}%
  \BibitemOpen
  \bibfield  {author} {\bibinfo {author} {\bibfnamefont {T.}~\bibnamefont
  {Fuchs}}, \bibinfo {author} {\bibfnamefont {J.}~\bibnamefont {Gegelia}},
  \bibinfo {author} {\bibfnamefont {G.}~\bibnamefont {Japaridze}},\ and\
  \bibinfo {author} {\bibfnamefont {S.}~\bibnamefont {Scherer}},\ }\href@noop
  {} {\bibfield  {journal} {\bibinfo  {journal} {Physical Review D}\ }\textbf
  {\bibinfo {volume} {68}},\ \bibinfo {pages} {056005} (\bibinfo {year}
  {2003})}\BibitemShut {NoStop}%
\bibitem [{\citenamefont {Schindler}\ \emph {et~al.}(2005)\citenamefont
  {Schindler}, \citenamefont {Gegelia},\ and\ \citenamefont
  {Scherer}}]{schindler2005}%
  \BibitemOpen
  \bibfield  {author} {\bibinfo {author} {\bibfnamefont {M.~R.}\ \bibnamefont
  {Schindler}}, \bibinfo {author} {\bibfnamefont {J.}~\bibnamefont {Gegelia}},\
  and\ \bibinfo {author} {\bibfnamefont {S.}~\bibnamefont {Scherer}},\ }\href
  {https://doi.org/10.1140/epja/i2005-10145-8} {\bibfield  {journal} {\bibinfo
  {journal} {European Physical Journal A}\ }\textbf {\bibinfo {volume} {26}},\
  \bibinfo {pages} {1} (\bibinfo {year} {2005})},\ \Eprint
  {https://arxiv.org/abs/nucl-th/0509005} {arXiv:nucl-th/0509005} \BibitemShut
  {NoStop}%
\bibitem [{\citenamefont {Kawarabayashi}\ and\ \citenamefont
  {Suzuki}(1966)}]{ksrf1}%
  \BibitemOpen
  \bibfield  {author} {\bibinfo {author} {\bibfnamefont {K.}~\bibnamefont
  {Kawarabayashi}}\ and\ \bibinfo {author} {\bibfnamefont {M.}~\bibnamefont
  {Suzuki}},\ }\href@noop {} {\bibfield  {journal} {\bibinfo  {journal}
  {Physical Review Letters}\ }\textbf {\bibinfo {volume} {16}},\ \bibinfo
  {pages} {255} (\bibinfo {year} {1966})}\BibitemShut {NoStop}%
\bibitem [{\citenamefont {Riazuddin}(1966)}]{ksrf2}%
  \BibitemOpen
  \bibfield  {author} {\bibinfo {author} {\bibfnamefont {F.}~\bibnamefont
  {Riazuddin}},\ }\href@noop {} {\bibfield  {journal} {\bibinfo  {journal}
  {Physical Review}\ }\textbf {\bibinfo {volume} {147}},\ \bibinfo {pages}
  {1071} (\bibinfo {year} {1966})}\BibitemShut {NoStop}%
\bibitem [{\citenamefont {Takahashi}\ \emph {et~al.}(2026)\citenamefont
  {Takahashi} \emph {et~al.}}]{pdg2026}%
  \BibitemOpen
  \bibfield  {author} {\bibinfo {author} {\bibfnamefont {F.}~\bibnamefont
  {Takahashi}} \emph {et~al.} (\bibinfo {collaboration} {Particle Data
  Group}),\ }\href {https://doi.org/10.1142/S0217751X26300115} {\bibfield
  {journal} {\bibinfo  {journal} {International Journal of Modern Physics A}\
  }\textbf {\bibinfo {volume} {41}},\ \bibinfo {pages} {2630011} (\bibinfo
  {year} {2026})}\BibitemShut {NoStop}%
\bibitem [{\citenamefont {Fujikawa}\ \emph {et~al.}(2008)\citenamefont
  {Fujikawa}, \citenamefont {Hayashii}, \citenamefont {Eidelman}, \citenamefont
  {Adachi}, \citenamefont {Aihara}, \citenamefont {Arinstein}, \citenamefont
  {Aulchenko}, \citenamefont {Aushev}, \citenamefont {Bakich}, \citenamefont
  {Balagura} \emph {et~al.}}]{Belle}%
  \BibitemOpen
  \bibfield  {author} {\bibinfo {author} {\bibfnamefont {M.}~\bibnamefont
  {Fujikawa}}, \bibinfo {author} {\bibfnamefont {H.}~\bibnamefont {Hayashii}},
  \bibinfo {author} {\bibfnamefont {S.}~\bibnamefont {Eidelman}}, \bibinfo
  {author} {\bibfnamefont {I.}~\bibnamefont {Adachi}}, \bibinfo {author}
  {\bibfnamefont {H.}~\bibnamefont {Aihara}}, \bibinfo {author} {\bibfnamefont
  {K.}~\bibnamefont {Arinstein}}, \bibinfo {author} {\bibfnamefont
  {V.}~\bibnamefont {Aulchenko}}, \bibinfo {author} {\bibfnamefont
  {T.}~\bibnamefont {Aushev}}, \bibinfo {author} {\bibfnamefont
  {A.}~\bibnamefont {Bakich}}, \bibinfo {author} {\bibfnamefont
  {V.}~\bibnamefont {Balagura}}, \emph {et~al.},\ }\href@noop {} {\bibfield
  {journal} {\bibinfo  {journal} {Physical Review D}\ }\textbf {\bibinfo
  {volume} {78}},\ \bibinfo {pages} {072006} (\bibinfo {year}
  {2008})}\BibitemShut {NoStop}%
\bibitem [{\citenamefont {Ambrosino}\ \emph {et~al.}(2011)\citenamefont
  {Ambrosino}, \citenamefont {Archilli}, \citenamefont {Beltrame},
  \citenamefont {Bencivenni}, \citenamefont {Bini}, \citenamefont {Bloise},
  \citenamefont {Bocchetta}, \citenamefont {Bossi}, \citenamefont {Branchini},
  \citenamefont {Capon} \emph {et~al.}}]{KLOE}%
  \BibitemOpen
  \bibfield  {author} {\bibinfo {author} {\bibfnamefont {F.}~\bibnamefont
  {Ambrosino}}, \bibinfo {author} {\bibfnamefont {F.}~\bibnamefont {Archilli}},
  \bibinfo {author} {\bibfnamefont {P.}~\bibnamefont {Beltrame}}, \bibinfo
  {author} {\bibfnamefont {G.}~\bibnamefont {Bencivenni}}, \bibinfo {author}
  {\bibfnamefont {C.}~\bibnamefont {Bini}}, \bibinfo {author} {\bibfnamefont
  {C.}~\bibnamefont {Bloise}}, \bibinfo {author} {\bibfnamefont
  {S.}~\bibnamefont {Bocchetta}}, \bibinfo {author} {\bibfnamefont
  {F.}~\bibnamefont {Bossi}}, \bibinfo {author} {\bibfnamefont
  {P.}~\bibnamefont {Branchini}}, \bibinfo {author} {\bibfnamefont
  {G.}~\bibnamefont {Capon}}, \emph {et~al.},\ }\href@noop {} {\bibfield
  {journal} {\bibinfo  {journal} {Physics Letters B}\ }\textbf {\bibinfo
  {volume} {700}},\ \bibinfo {pages} {102} (\bibinfo {year}
  {2011})}\BibitemShut {NoStop}%
\bibitem [{\citenamefont {Achasov}\ \emph {et~al.}(2006)\citenamefont
  {Achasov}, \citenamefont {Beloborodov}, \citenamefont {Berdyugin},
  \citenamefont {Bogdanchikov}, \citenamefont {Bozhenok}, \citenamefont
  {Bukin}, \citenamefont {Bukin}, \citenamefont {Dimova}, \citenamefont
  {Druzhinin}, \citenamefont {Golubev} \emph {et~al.}}]{SND}%
  \BibitemOpen
  \bibfield  {author} {\bibinfo {author} {\bibfnamefont {M.}~\bibnamefont
  {Achasov}}, \bibinfo {author} {\bibfnamefont {K.}~\bibnamefont
  {Beloborodov}}, \bibinfo {author} {\bibfnamefont {A.}~\bibnamefont
  {Berdyugin}}, \bibinfo {author} {\bibfnamefont {A.}~\bibnamefont
  {Bogdanchikov}}, \bibinfo {author} {\bibfnamefont {A.}~\bibnamefont
  {Bozhenok}}, \bibinfo {author} {\bibfnamefont {A.}~\bibnamefont {Bukin}},
  \bibinfo {author} {\bibfnamefont {D.}~\bibnamefont {Bukin}}, \bibinfo
  {author} {\bibfnamefont {T.}~\bibnamefont {Dimova}}, \bibinfo {author}
  {\bibfnamefont {V.}~\bibnamefont {Druzhinin}}, \bibinfo {author}
  {\bibfnamefont {V.}~\bibnamefont {Golubev}}, \emph {et~al.},\ }\href@noop {}
  {\bibfield  {journal} {\bibinfo  {journal} {Journal of Experimental and
  Theoretical Physics}\ }\textbf {\bibinfo {volume} {103}},\ \bibinfo {pages}
  {380} (\bibinfo {year} {2006})}\BibitemShut {NoStop}%
\bibitem [{\citenamefont {Pascalutsa}\ and\ \citenamefont
  {Phillips}(2003)}]{pascalutsa}%
  \BibitemOpen
  \bibfield  {author} {\bibinfo {author} {\bibfnamefont {V.}~\bibnamefont
  {Pascalutsa}}\ and\ \bibinfo {author} {\bibfnamefont {D.~R.}\ \bibnamefont
  {Phillips}},\ }\href@noop {} {\bibfield  {journal} {\bibinfo  {journal}
  {Physical Review C}\ }\textbf {\bibinfo {volume} {67}},\ \bibinfo {pages}
  {055202} (\bibinfo {year} {2003})}\BibitemShut {NoStop}%
\bibitem [{\citenamefont {Lees}\ \emph {et~al.}(2012)\citenamefont {Lees} \emph
  {et~al.}}]{BaBar2012}%
  \BibitemOpen
  \bibfield  {author} {\bibinfo {author} {\bibfnamefont {J.~P.}\ \bibnamefont
  {Lees}} \emph {et~al.} (\bibinfo {collaboration} {BABAR Collaboration}),\
  }\href {https://doi.org/10.1103/PhysRevD.86.032013} {\bibfield  {journal}
  {\bibinfo  {journal} {Physical Review D}\ }\textbf {\bibinfo {volume} {86}},\
  \bibinfo {pages} {032013} (\bibinfo {year} {2012})},\ \Eprint
  {https://arxiv.org/abs/1205.2228} {arXiv:1205.2228} \BibitemShut {NoStop}%
\bibitem [{\citenamefont {Ablikim}\ \emph {et~al.}(2016)\citenamefont {Ablikim}
  \emph {et~al.}}]{BESIII2016}%
  \BibitemOpen
  \bibfield  {author} {\bibinfo {author} {\bibfnamefont {M.}~\bibnamefont
  {Ablikim}} \emph {et~al.} (\bibinfo {collaboration} {BESIII Collaboration}),\
  }\href {https://doi.org/10.1016/j.physletb.2015.11.043} {\bibfield  {journal}
  {\bibinfo  {journal} {Physics Letters B}\ }\textbf {\bibinfo {volume}
  {753}},\ \bibinfo {pages} {629} (\bibinfo {year} {2016})},\ \bibinfo {note}
  {corrigendum: Phys. Lett. B 812, 135982 (2021)}\BibitemShut {NoStop}%
\bibitem [{\citenamefont {Ignatov}\ \emph {et~al.}(2024)\citenamefont {Ignatov}
  \emph {et~al.}}]{CMD32024}%
  \BibitemOpen
  \bibfield  {author} {\bibinfo {author} {\bibfnamefont {F.~V.}\ \bibnamefont
  {Ignatov}} \emph {et~al.} (\bibinfo {collaboration} {CMD-3 Collaboration}),\
  }\href {https://doi.org/10.1103/PhysRevLett.132.231903} {\bibfield  {journal}
  {\bibinfo  {journal} {Physical Review Letters}\ }\textbf {\bibinfo {volume}
  {132}},\ \bibinfo {pages} {231903} (\bibinfo {year} {2024})}\BibitemShut
  {NoStop}%
\bibitem [{\citenamefont {Davier}\ \emph {et~al.}(2024)\citenamefont {Davier},
  \citenamefont {Hoecker}, \citenamefont {Lutz}, \citenamefont {Malaescu},\
  and\ \citenamefont {Zhang}}]{Davier2024}%
  \BibitemOpen
  \bibfield  {author} {\bibinfo {author} {\bibfnamefont {M.}~\bibnamefont
  {Davier}}, \bibinfo {author} {\bibfnamefont {A.}~\bibnamefont {Hoecker}},
  \bibinfo {author} {\bibfnamefont {A.-M.}\ \bibnamefont {Lutz}}, \bibinfo
  {author} {\bibfnamefont {B.}~\bibnamefont {Malaescu}},\ and\ \bibinfo
  {author} {\bibfnamefont {Z.}~\bibnamefont {Zhang}},\ }\href
  {https://doi.org/10.1140/epjc/s10052-024-12964-7} {\bibfield  {journal}
  {\bibinfo  {journal} {European Physical Journal C}\ }\textbf {\bibinfo
  {volume} {84}},\ \bibinfo {pages} {721} (\bibinfo {year} {2024})},\ \Eprint
  {https://arxiv.org/abs/2312.02053} {arXiv:2312.02053} \BibitemShut {NoStop}%
\bibitem [{\citenamefont {O'Connell}\ \emph {et~al.}(1997)\citenamefont
  {O'Connell}, \citenamefont {Pearce}, \citenamefont {Thomas},\ and\
  \citenamefont {Williams}}]{thomas}%
  \BibitemOpen
  \bibfield  {author} {\bibinfo {author} {\bibfnamefont {H.~B.}\ \bibnamefont
  {O'Connell}}, \bibinfo {author} {\bibfnamefont {B.}~\bibnamefont {Pearce}},
  \bibinfo {author} {\bibfnamefont {A.~W.}\ \bibnamefont {Thomas}},\ and\
  \bibinfo {author} {\bibfnamefont {A.~G.}\ \bibnamefont {Williams}},\
  }\href@noop {} {\bibfield  {journal} {\bibinfo  {journal} {Progress in
  Particle and Nuclear Physics}\ }\textbf {\bibinfo {volume} {39}},\ \bibinfo
  {pages} {201} (\bibinfo {year} {1997})}\BibitemShut {NoStop}%
\bibitem [{\citenamefont {Urech}(1995)}]{urec}%
  \BibitemOpen
  \bibfield  {author} {\bibinfo {author} {\bibfnamefont {R.}~\bibnamefont
  {Urech}},\ }\href@noop {} {\bibfield  {journal} {\bibinfo  {journal} {Physics
  Letters B}\ }\textbf {\bibinfo {volume} {355}},\ \bibinfo {pages} {308}
  (\bibinfo {year} {1995})}\BibitemShut {NoStop}%
\bibitem [{\citenamefont {Gasser}\ and\ \citenamefont
  {Leutwyler}(1982)}]{qmass}%
  \BibitemOpen
  \bibfield  {author} {\bibinfo {author} {\bibfnamefont {J.}~\bibnamefont
  {Gasser}}\ and\ \bibinfo {author} {\bibfnamefont {H.}~\bibnamefont
  {Leutwyler}},\ }\href@noop {} {\bibfield  {journal} {\bibinfo  {journal}
  {Physics Reports}\ }\textbf {\bibinfo {volume} {87}},\ \bibinfo {pages} {77}
  (\bibinfo {year} {1982})}\BibitemShut {NoStop}%
\bibitem [{\citenamefont {Park}(2026)}]{ParkNN2026}%
  \BibitemOpen
  \bibfield  {author} {\bibinfo {author} {\bibfnamefont {T.-S.}\ \bibnamefont
  {Park}}} (\bibinfo {year} {2026}),\ \bibinfo {note} {companion paper, in
  preparation}\BibitemShut {NoStop}%
\end{thebibliography}
%

\end{document}